\documentclass[12pt]{article}

\usepackage{amssymb}
\usepackage{amsthm}

\usepackage[latin1]{inputenc}

\usepackage{logique}

\newcommand{\NN}{\mathbb{N}}
\newcommand{\CC}{\mathbb{C}}

\newcommand{\HH}{\mbox{\sffamily H}}
\newcommand{\PP}{\mbox{\footnotesize\sffamily P}}
\newcommand{\indi}{^{\mbox{\footnotesize int}}}

\newcommand{\nnforce}{\mbox{ $\;\not\!\!\!\!|\!|\!\!-\,$}}

\renewcommand{\gmg}{``~}
\renewcommand{\gmd}{~''}
\renewcommand{\gmdd}{~''}

\newtheorem{theorem}{Théorème}
\newtheorem{lemma}[theorem]{Lemme}

\newtheorem{proposition}[theorem]{Proposition}

\author{Jean-Louis Krivine\\
\footnotesize{Université Paris VII, C.N.R.S.}\\
}

\title{Structures de réalisabilité, RAM et ultrafiltre sur $\NN$}
\date{\footnotesize {8 septembre 2008}}

\begin{document}
\maketitle

\section*{Introduction}
On montre ici comment transformer en programmes les démonstrations utilisant l'axiome du choix dépendant et
l'existence d'un \emph{ultrafiltre non trivial sur $\NN$} (qu'on peut, de plus, supposer \emph{sélectif}
\footnote{Un ultrafiltre ${\cal U}$ sur $\NN$ est dit \emph{sélectif} (voir [1]) si, pour toute relation
d'équivalence sur $\NN$, dont aucune classe n'est dans ${\cal U}$, il existe un élément de ${\cal U}$ qui
choisit un élément de chaque classe.}).
La méthode est une extension de la technique de \emph{réalisabilité en logique classique} introduite
dans [3,4,5], et de la méthode du \emph{forcing} bien connue en théorie des ensembles.
Pour simplifier un peu l'exposé, on prend comme cadre axiomatique, l'arithmé\-tique du second ordre classique
avec axi\-ome du choix dépendant (qu'on appelle aussi l'\emph{Ana\-lyse}). En suivant les idées développées
dans~[2], on peut étendre ce résultat à la théorie ZF (avec axiome du choix dépendant) en ajoutant
l'axiome de l'ultrafiltre~: \gmg il existe un ultra\-filtre sur toute algèbre de Boole\gmd. On le fera dans un
prochain article, où on traitera également des axiomes comme~:
\gmg il existe un bon ordre sur ${\cal P}(\NN)$ dont tout segment initial est dénombrable\gmd, ou
\gmg il existe un bon ordre sur ${\cal P}({\cal P}(\NN))$\gmd.\\
En fait, il est clair que la même méthode permet de traiter l'axiome du bon ordre (c'est-à-dire l'axiome du
choix). Mais, à ce jour, je n'en ai pas rédigé la démonstration.

\smallskip\noindent
La signification informatique de ces résultats est intéressante~: on a vu, dans [3,4,5], que l'on pouvait interpréter
l'axiome du choix dépendant en introduisant une nouvelle instruction de {\em signature} ou d'\emph{horloge}.
Pour interpréter l'axiome d'ultrafiltre ou de bon ordre, il faut maintenant ajou\-ter des instructions de \emph{lecture} et d'\emph{écriture} dans une \emph{mémoire globale} (appelée aussi \gmg tas\gmdd ou
\gmg Random Access Memory\gmdd en informatique).\\
Mais la manipulation logique de ces instructions n'est possible qu'au prix d'une extension de la réalisabilité classique~; c'est pourquoi on introduit la notion générale de \emph{structure de réali\-sabilité}.\\
Il est remarquable que les axiomes pour les conditions de forcing (semi-treillis) s'inter\-prè\-tent ici comme les indispensables \emph{programmes de gestion de la mémoire}. Yves Legrand\-gérard a fait le rapprochement avec les instructions telles que \texttt{malloc} et \texttt{free} dans le langage~C. Or, la propriété essentielle de la
mémoire globale est d'être indépendante de la pile, c'est-à-dire qu'elle n'est pas mémorisée par l'ins\-truction $\ccc$ associée au raisonnement par l'absurde.\\
Il est amusant de noter que cela explique, a posteriori, une constatation empirique bien connue des
théo\-riciens des ensembles, qui est le \gmg parfum intuitionniste\gmdd du forcing.

\smallskip\noindent
Je remercie Yves Legrandgérard, dont les connaissances approfondies en programmation système et réseau
ont été capitales pour l'appréhension du sens informatique des résultats énoncés ici.\\
Nous comptons développer ensemble ce sujet dans un prochain article.

\section*{La sémantique des programmes}

\subsection*{Structures de réalisabilité}
Une {\em quasi-preuve} est un $\lbd$-terme $t[x_1,\ldots,x_k]$ avec la constante
$\ccc$. Les variables libres de $t[x_1,\ldots,x_k]$ se trouvent parmi
$x_1,\ldots,x_k$. L'ensemble des quasi-preuves est noté \QP, l'en\-semble des quasi-preuves
closes est noté $\QP_0$.\\
Toute démonstration, en Analyse ou dans ZF\/, de \ $x_1{:}A_1,\ldots,x_k{:}A_k\vdash t{:}A$
donne une quasi-preuve $t[x_1,\ldots,x_k]$~(voir ci-dessous les règles de démonstration et de typage
pour la logique classique du second ordre).

\smallskip\noindent
Une {\em structure de réalisabilité} est la donnée de trois ensembles $\LLbd,\PPi,\LLbd\star\PPi$
(termes, piles et processus), avec les opérations suivantes~:

\smallskip\noindent
$\bullet$~~Une application $(\xi,\pi)\mapsto\xi\ps\pi$ de $\LLbd\fois\PPi$ dans $\PPi$
({\em empiler}).\\
$\bullet$~~Une application $(\xi,\pi)\mapsto\xi\star\pi$ de $\LLbd\fois\PPi$ dans $\LLbd\star\PPi$
({\em processus}).\\
$\bullet$~~Une application $\pi\mapsto\kk_\pi$ de $\PPi$ dans $\LLbd$ ({\em continuation}).\\
Une \emph{substitution}, notée $[\xi_1/x_1,\ldots,\xi_k/x_k]$ est,
par définition, un ensemble fini de couples $\{(x_1,\xi_1),\ldots,(x_k,\xi_k)\}$, où $x_1,\ldots,x_k$
sont des variables \emph{distinctes} et $\xi_1,\ldots,\xi_k\in\LLbd$.\\
On fixe un ensemble ${\cal S}$ de substitutions qui contient la substitution vide (notée $[]$) et tel que
si $[\xi/x,\xi_1/x_1,\ldots,\xi_k/x_k]\in{\cal S}$, alors $[\xi_1/x_1,\ldots,\xi_k/x_k]\in{\cal S}$.\\
$\bullet$~~Pour chaque quasi-preuve $t[x_1,\ldots,x_k]$ et chaque substitution
$[\xi_1/x_1,\ldots,\xi_k/x_k]\in{\cal S}$, on se donne $t[\xi_1/x_1,\ldots,\xi_k/x_k]\in\LLbd$.

\smallskip\noindent
{\small{\bfseries Remarque.} Dans la plupart des structures de réalisabilité considérées ici, on pourra
définir une opération binaire sur $\LLbd$, analogue à \emph{l'application} dans les $\lbd$-termes. Mais
les axiomes de structure de réalisabilité ne comportent pas d'opération binaire sur $\LLbd$.}

\smallskip\noindent
On se donne un ensemble $\bbot$ de processus et on définit sur $\LLbd$ une relation de préordre en posant~: \
$\xi\le\eta$ $\Dbfl$ $(\pt\pi\in\PPi)(\eta\star\pi\in\bbot$ $\Fl$ $\xi\star\pi\in\bbot$).\\
On suppose l'ensemble $\bbot$ \emph{saturé}, ce qui veut dire qu'il a les propriétés
suivantes~:\label{compatibilite}\\
1. Si $[\xi_1/x_1,\ldots,\xi_k/x_k]\in{\cal S}$ et si \ $\xi_i\star\pi\in\bbot$, alors $x_i[\xi_1/x_1,\ldots,\xi_k/x_k]\star\pi\in\bbot$~;\\
autrement dit \ $x_i[\xi_1/x_1,\ldots,\xi_k/x_k]\le\xi_i$.\\
2. Si $[\xi_1/x_1,\ldots,\xi_k/x_k]\in{\cal S}$ et si, pour tous $\xi'_1\le\xi_1,\ldots,\xi'_k\le\xi_k$
avec ${[\xi'_1/x_1,\ldots,\xi'_k/x_k]\in{\cal S}}$, on a
$t[\xi'_1/x_1,\ldots,\xi'_k/x_k]\star u[\xi'_1/x_1,\ldots,\xi'_k/x_k]\ps\pi\in\bbot$,
alors $tu[\xi_1/x_1,\ldots,\xi_k/x_k]\star\pi\in\bbot$.\\
3. Si $[\xi_1/x_1,\ldots,\xi_k/x_k]\in{\cal S}$ et si, pour tous
$\xi'\le\xi,\xi'_1\le\xi_1,\ldots,\xi'_k\le\xi_k$\\
avec ${[\xi'/x,\xi'_1/x_1,\ldots,\xi'_k/x_k]\in{\cal S}}$, on a \
$t[\xi'/x,\xi'_1/x_1,\ldots,\xi'_k/x_k]\star\pi\in\bbot$\\
alors $\lbd x\,t[\xi_1/x_1,\ldots,\xi_k/x_k]\star\xi\ps\pi\in\bbot$.\\
4. Si $[\xi_1/x_1,\ldots,\xi_k/x_k]\in{\cal S}$, et si $\xi\star\kk_\pi\ps\pi\in\bbot$ \ alors \ $\ccc[\xi_1/x_1,\ldots,\xi_k/x_k]\star\xi\ps\pi\in\bbot$.\\
5. $\xi\star\pi\in\bbot$ $\Fl$ $\kk_\pi\star\xi\ps\pi'\in\bbot$.

\smallskip\noindent
En itérant la condition~3, on obtient~:\\
3. Si $[\xi_1/x_1,\ldots,\xi_k/x_k]\in{\cal S}$ et si quels que soient
$\xi'_1\le\xi_1,\ldots,\xi'_k\le\xi_k,\eta'_1\le\eta_1,\ldots,\eta'_l\le\eta_l$\\
avec $[\xi'_1/x_1,\ldots,\xi'_k/x_k,\eta'_1/y_1,\ldots,\eta'_l/y_l]\in{\cal S}$, on a \
${t[\xi'_1/x_1,\ldots,\xi'_k/x_k,\eta'_1/y_1,\ldots,\eta'_l/y_l]\star\pi\in\bbot}$\\
alors $\lbd y_1\ldots\lbd y_l\,t[\xi_1/x_1,\ldots,\xi_k/x_k]\star\eta_1\ps\ldots\ps\eta_l\ps\pi\in\bbot$.

\subsection*{Réalisabilité en arithmétique du second ordre}
Les formules considérées sont écrites en logique du second ordre, avec $\pt,\to$ et $\bot$ comme seuls
symboles logiques. Les variables d'individu sont notées $x,y,\ldots,$ et les variables de
prédicats~$X,Y,\ldots$ \ Chaque variable de prédicat (on dira aussi variable de relation) a une \emph{arité}
qui est un entier. Une variable d'arité 0 est appelée \emph{variable propositionnelle}.\\
A chaque fonction $f:\NN^k\to\NN$ est associé un symbole de fonction d'arité~$k$, noté également~$f$.\\
On utilisera la notation \ $A_1,A_2,\ldots,A_k\to A$ pour désigner la formule~:\\
$A_1\to(A_2\to(\cdots\to(A_k\to A)))$.\\
{\em Un paramètre du second ordre d'arité $k$} est une application ${\cal X}:\ennl^k\to{\cal P}(\PPi)$.\\
Une {\em interprétation} $\cal I$ est une application qui associe un individu (entier)
à chaque variable d'individu et un paramètre d'arité $k$ à chaque variable du second
ordre d'arité $k$.\\
${\cal I}[x\lf n]$ (resp. ${\cal I}[X\lf{\cal X}]$) est l'interprétation obtenue en
changeant, dans ${\cal I}$, la valeur de la variable $x$ (resp. $X$) et en lui donnant
la valeur $n$ (resp. ${\cal X}$).\\
Pour toute formule $A$, on désigne par $A^{\cal I}$ la {\em formule close avec paramètres}
obtenue en remplaçant, dans $A$, chaque variable libre par la valeur donnée par
$\cal I$.\\
On définit $\|A^{\cal I}\|\subset\PPi$ pour toute formule $A^{\cal I}$
close avec paramètres~; c'est la {\em valeur de vérité de la formule $A^{\cal I}$}.\\
Pour $\xi\in\LLbd$, $\xi\force A^{\cal I}$ (lire \gmg $\xi$ réalise $A^{\cal I}$\gmd) est, par définition,
$(\pt\pi\in\|A^{\cal I}\|)\,\xi\star\pi\in\bbot$.

\smallskip\noindent
La définition de $\|A^{\cal I}\|$ se fait par récurrence sur la longueur de la formule $A$~:

\smallskip\noindent
$\bullet$~~Si $A$ est atomique, on a $A^{\cal I}\equiv\|{\cal X}(t_1,\ldots,t_k)\|$, où ${\cal X}$ est un paramètre
d'arité $k$ et $t_1,\ldots,t_k$ des termes clos d'individu, qui ont donc respectivement les valeurs
${n_1,\ldots,n_k\in\NN}$. On pose alors $\|A^{\cal I}\|={\cal X}(n_1,\ldots,n_k)$.\\
$\bullet$~~$\|(A\to B)^{\cal I}\|$ est, par définition, $\{\xi\ps\pi;$
$\xi\force A^{\cal I},\pi\in\|B^{\cal I}\|\}$~;\\
$\bullet$~~$\|(\pt x\,A)^{\cal I}\|$ est, par définition,
$\bigcup_{n\in\NN}\| A^{{\cal I}[x\lf n]}\|$.\\
$\bullet$~~$\|(\pt X\,A)^{\cal I}\|$ est, par définition,
$\bigcup\{\| A^{{\cal I}[X\lf{\cal X}]}\|;$ ${\cal X}\in{\cal P}(\PPi)^{\ennl^k}\}$.

\smallskip\noindent
{\small{\bfseries Remarques.}\\
i) Si $\xi\force A^{\cal I}$ et $\xi'\le\xi$, alors $\xi'\force A^{\cal I}$.\\
ii) On utilisera les notations $\vv A^{\cal I}\vv$ et $\xi\fforce A^{\cal I}$ si on a déjà
défini une autre structure de réalisabilité.}

\smallskip\noindent
On donne ci-dessous les règles de démonstration et de typage en logique classique du se\-cond ordre
(voir [3,4,5])~; dans ces règles, $\Gamma$ désigne un {\em contexte}, c'est-à-dire une expression
de la forme $x_1:A_1,\,\ldots,\,x_n:A_n$. Dans une expresssion comme \gmg$t:A$\gmd, $t$ est
une quasi-preuve et $A$ une formule du second ordre.

\smallskip\noindent
1.~$\Gamma\vdash x_i:A_i$ ($1\le i\le n$)\hspace{\fill}(Axiome)\\
2.~$\Gamma\vdash t:A\to B$, $\,\Gamma\vdash u:A$ $\;\Fl\;$
$\Gamma\vdash tu:B$\hspace{\fill}(Modus ponens)\\
3. $\Gamma, x:A\vdash t:B$ $\;\Fl\;$ $\Gamma\vdash\lambda x\,t:A\to B$\hspace{\fill}(Introduction de $\to$)\\
4. $\Gamma\vdash\ccc:((A\to B)\to A)\to A$\hspace{\fill}(Loi de Peirce)\\
5. $\Gamma\vdash t:A$ $\;\Fl\;$ $\Gamma\vdash t:\forall x\,A$ (resp.
$\forall X\,A$)\hspace{\fill}(Introduction de $\pt$)\\
si $x$ (resp. $X$) n'est pas libre dans $\Gamma$.\\
6. $\Gamma\vdash t:\forall x\,A$ $\Fl$ $\Gamma\vdash t:A[\tau/x]$\ \
pour tout terme d'individu $\tau$\hspace{\fill}(Elimination de $\pt x$)\\
7. $\Gamma\vdash t:\forall X\,A$ $\Fl$ $\Gamma\vdash t:A[F/Xx_1\ldots x_k]$ pour toute formule $F$
\hspace{\fill}(Elimination de $\pt X$)

\begin{theorem}[Lemme d'adéquation]\label{adequat_gen}\ \\
Si \ $x_1:A_1,\ldots,x_k:A_k\vdash t:A$, si $[\xi_1/x_1,\ldots,\xi_k/x_k]\in{\cal S}$ et si \
$\xi_i\force A_i^{\cal I}$ $(1\le i\le k)$,\\
alors $t[\xi_1/x_1,\ldots,\xi_k/x_k]\force A^{\cal I}$.\\
En particulier, si $A$ est close et si \ $\vdash t:A$, alors $t[\xi_1/x_1,\ldots,\xi_k/x_k]\force A$ \
pour toute substitution \ $[\xi_1/x_1,\ldots,\xi_k/x_k]\in{\cal S}$. On écrira cette propriété, en abrégé, \
$t[{\cal S}]\force A$ ou même $t\force A$.
\end{theorem}\noindent
Démonstration par récurrence sur la longueur de la preuve de
$x_1:A_1,\ldots,x_k:A_k\vdash t:A$. On considère la dernière règle utilisée.

\smallskip\noindent
1. Axiome.  On a $\xi_i\force A_i^{\cal I}$ et donc $x_i[\xi_1/x_1,\ldots,\xi_k/x_k]\force A_i^{\cal I}$,
d'après l'hypothèse~1 sur $\bbot$

\smallskip\noindent
2. Modus ponens. On a $t=uv$~; $x_1:A_1,\ldots,x_k:A_k\vdash u:B\to A$ et $v:B$.\\
Etant donnée $\pi\in\|A^{\cal I}\|$, on doit montrer \ $(uv)[\xi_1/x_1,\ldots,\xi_k/x_k]\star\pi\in\bbot$.\\
D'après l'hypothèse~2 sur $\bbot$, il suffit de montrer~:\\
\centerline{$u[\xi'_1/x_1,\ldots,\xi'_k/x_k]\star v[\xi'_1/x_1,\ldots,\xi'_k/x_k]\ps\pi\in\bbot$}

\noindent
pour $\xi'_1\le\xi_1,\ldots,\xi'_k\le\xi_k$ avec $[\xi'_1/x_1,\ldots,\xi'_k/x_k]\in{\cal S}$.\\
Or, on a \ $\xi_i\force A_i^{\cal I}$ et donc  $\xi'_i\force A_i^{\cal I}$, puisque $\xi'_i\le\xi_i$.
Par hypothèse de récurrence, on a~:\\
$v[\xi'_1/x_1,\ldots,\xi'_k/x_k]\force B^{\cal I}$ et par suite, \
$v[\xi'_1/x_1,\ldots,\xi'_k/x_k]\ps\pi\in\|B^{\cal I}\to A^{\cal I}\|$.\\
Or, par hypothèse de récurrence, on a aussi $u[\xi'_1/x_1,\ldots,\xi'_k/x_k]\force B^{\cal I}\to A^{\cal I}$,
d'où le résultat.

\smallskip\noindent
3. Introduction de $\to$. On a $A=B\to C$, $t=\lbd x\,u$. On doit montrer~:\\
$\lbd x\,u[\xi_1/x_1,\ldots,\xi_k/x_k]\force B^{\cal I}\to C^{\cal I}$
et on considère donc $\xi\force B^{\cal I}$, $\pi\in\| C^{\cal I}\|$. On est ramené à
montrer $\lbd x\,u[\xi_1/x_1,\ldots,\xi_k/x_k]\star\xi\ps\pi\in\bbot$. Pour cela,
d'après l'hypothèse~3 sur $\bbot$, il suffit de montrer
$u[\xi'/x,\xi'_1/x_1,\ldots,\xi'_k/x_k]\star\pi\in\bbot$, quels que soient
$\xi'\le\xi,\xi'_1\le\xi_1,\ldots,\xi'_k\le\xi_k$, tels que $[\xi'/x,\xi'_1/x_1,\ldots,\xi'_k/x_k]\in{\cal S}$.\\
Or, on a $\xi\force B^{\cal I}$ et donc  $\xi'\force B^{\cal I}$, puisque $\xi'\le\xi$. De même,
$\xi'_i\force A_i^{\cal I}$~; d'après l'hypothèse de récurrence, on a donc
$u[\xi'/x,\xi'_1/x_1,\ldots,\xi'_k/x_k]\force C^{\cal I}$, d'où le résultat, puisque
${\pi\in\| C^{\cal I}\|}$.

\smallskip\noindent
4. Loi de Peirce. On montre d'abord~:
\begin{lemma}\label{continuation}
Soit ${\cal A}\subset\PPi$ une valeur de vérité. Si $\pi\in{\cal A}$, \ alors \ $\kk_\pi\force\neg{\cal A}$.
\end{lemma}\noindent
Soient $\xi\force{\cal A}$ et $\rho\in\PPi$~; on doit montrer $\kk_\pi\star\xi\ps\rho\in\bbot$, soit
$\xi\star\pi\in\bbot$, ce qui est clair.

\cqfd

\smallskip\noindent
On doit montrer que $\ccc[\xi_1/x_1,\ldots,\xi_k/x_k]\force(\neg A\to A)\to A$.\\
Soient donc \ $\xi\force\neg A\to A$ et $\pi\in\|A\|$. On doit montrer que \
${\ccc[\xi_1/x_1,\ldots,\xi_k/x_k]\star\xi\ps\pi\in\bbot}$, soit $\xi\star\kk_\pi\ps\pi\in\bbot$.
D'après l'hypothèse sur $\xi$ et $\pi$, il suffit de montrer que $\kk_\pi\force\neg A$, ce qui résulte
du lemme~\ref{continuation}.

\smallskip\noindent
5. Introduction de $\pt$. On a $A=\pt X\,B$, $X$ n'étant pas libre dans $A_i$.
On doit montrer~:\\
$t[\xi_1/x_1,\ldots,\xi_k/x_k]\force(\pt X\,B)^{\cal I}$, c'est-à-dire
$t[\xi_1/x_1,\ldots,\xi_k/x_k]\force B^{\cal J}$ avec
${\cal J}={\cal I}[X\leftarrow{\cal X}]$.
Or on a, par hypothèse, $\xi_i\force A_i^{\cal I}$  donc $\xi_i\force A_i^{\cal J}$~:
en effet, comme $X$ n'est pas libre dans $A_i$, on a $\|A_i^{\cal I}\|=\|A_i^{\cal J}\|$.
L\ 'hypothèse de récurrence donne alors le résultat.

\smallskip\noindent
7. Elimination de $\pt X$. On a $A=B[F/Xx_1\ldots x_n]$ et on doit montrer~:\\
$t[\xi_1/x_1,\ldots,\xi_k/x_k]\force B[F/Xx_1\ldots x_n]^{\cal I}$
avec l'hypothèse $t[\xi_1/x_1,\ldots,\xi_k/x_k]\force(\pt X\,B)^{\cal I}$.\\
Cela découle du~:
\begin{lemma}
$\|B[F/Xx_1\ldots x_n]^{\cal I}\|=\|B\|^{{\cal I}[X\leftarrow{\cal X}]}$ où
${\cal X}:\ennl^n\to{\cal P}(\PPi)$ est défini par~:\\
${\cal X}(k_1,\ldots,k_n)=\| F^{{\cal I}[x_1\leftarrow k_1,\ldots,x_n\leftarrow k_n]}\|$.
\end{lemma}\noindent
Preuve par récurrence sur $B$. C'est trivial si $X$ n'est pas libre dans $B$.
Le seul cas intéressant de la récurrence est $B=\pt Y\,C$, et on a donc $Y\ne X$. On a alors~:\\
$\| B[F/Xx_1\ldots x_n]^{\cal I}\|=\|(\pt Y\,C[F/Xx_1\ldots x_n])^{\cal I}\|=
\bigcup_{\cal Y}\| C[F/Xx_1\ldots x_n]^{{\cal I}[Y\leftarrow{\cal Y}]}\|$.\\
Par hypothèse de récurrence, cela donne
$\bigcup_{\cal Y}\| C^{{\cal I}[Y\leftarrow{\cal Y}][X\leftarrow{\cal X}]}\|$, soit
$\bigcup_{\cal Y}\| C^{{\cal I}[X\leftarrow{\cal X}][Y\leftarrow{\cal Y}]}\|$
c'est-à-dire $\|(\pt Y\,C)^{{\cal I}[X\leftarrow{\cal X}]}\|$.

\cqfd

\smallskip\noindent
{\bfseries Définitions.}\\
i) La formule $x=y$ est, par définition, $\pt X(Xx\to Xy)$.\\
ii) Soit ${\cal E}\subset\LLbd$ et ${\cal X}\subset\PPi$ une valeur de vérité. On pose
${\cal E}\to{\cal X}=\{\xi\ps\pi;$ $\xi\in{\cal E},\pi\in{\cal X}\}$.\\
Cela permet de donner une valeur de vérité à des \gmg formules généralisées\gmd, du genre~:\\
$\pt x\pt X[{\cal E}(x,X)\to F(x,X)]$.\\
iii) Si $t\in\QP_0$, on désignera par $\{t[{\cal S}]\}$ l'ensemble $\{t[\xi_1/x_1,\ldots,\xi_k/x_k];$
${[\xi_1/x_1,\ldots,\xi_k/x_k]\in{\cal S}\}}$.\\
iv) Certains éléments de $\LLbd$ sont des \emph{instructions}. On impose alors à $\bbot$ des conditions
supplémentaires, qu'on appelle des \emph{règles de réduction}, notées $\xi\star\pi\succ\xi'\star\pi'$,
ce veut dire~: \ \
si $\xi'\star\pi'\in\bbot$, alors $\xi\star\pi\in\bbot$.\\
Un premier exemple est donné dans le théorème~\ref{memoire_gen}.

\begin{theorem}\label{neq}
On définit le prédicat binaire $\ne$ par $\|m\ne n\|=\PPi$ si $m=n$ et $\vide$ sinon. Alors~:\\
i) $\lbd x\,xI[{\cal S}]\force\pt x\pt y[(x=y\to\bot)\to x\ne y]$~;\\
ii) $\lbd x\lbd y\,yx[{\cal S}]\force\pt x\pt y[x\ne y\to(x=y\to\bot)]$.
\end{theorem}\noindent
i) Il suffit de montrer $\lbd x\,xI[{\cal S}]\force(m=m\to\bot)\to\bot$, ce qui découle de~:\\
$\vdash\lbd x\,xI:(m=m\to\bot)\to\bot$ et du théorème~\ref{adequat_gen} (lemme d'adéquation).\\
ii) On doit montrer $\lbd x\lbd y\,yx[{\cal S}]\force\bot,m=m\to\bot$ et $\top,(\top\to\bot)\to\bot$
ce qui découle de~:\\
$\vdash\lbd x\lbd y\,yx:\bot,m=m\to\bot$, \ $\vdash\lbd x\lbd y\,yx:\top,(\top\to\bot)\to\bot$
et du lemme d'adéquation.

\cqfd

\begin{theorem}\label{memoire_gen}
Soient $U=(u_n)_{n\in\NN}$ une suite d'éléments de $\LLbd$, et $T_U,S_U\in\LLbd$ deux instructions avec les règles de réduction~:\\
$T_U\star\phi\ps\nu\ps\pi\succ\nu\star S_U\ps\phi\ps u_0\ps\pi$~; \
$S_U\star\psi\ps u_n\ps\pi\succ\psi\star u_{n+1}\ps\pi$.	 Alors~:\\
$T_U\force\pt n\pt X[(\{u_n\}\to X),$int$(n)\to X]$.
\end{theorem}\noindent
Soient $n\in\mathbb{N}$, \ $\phi\force\{u_n\}\to X$, \ $\nu\force$ int$(n)$ et $\pi\in X$. On doit montrer
${T_U\star\phi\ps\nu\ps\pi\in\bbot}$, c'est-à-dire $\nu\star S_U\ps\phi\ps u_0\ps\pi\in\bbot$, d'après la
règle de réduction de $T_U$. Comme $\nu\force\mbox{int}(n)$, il suffit de trouver un prédicat unaire $Y$
tel que $S_U\force\pt y(Yy\to Ysy)$, $\phi\force Y0$ \ et ${u_0\ps\pi\in\|Yn\|}$.\\
On pose $Yi=\{u_{n-i}\ps\pi\}$ pour $0\le i\le n$ et $\|Yi\|=\vide$ pour $i>n$.\\
On a $u_0\ps\pi\in\|Yn\|$ par définition de $Yn$ et $\phi\force Y0$, par hypothèse sur $\phi$.\\
On montre $S_U\force Yi\to Ysi$~: c'est évident pour $i\ge n$, puisqu'alors $\|Ysi\|=\vide$.\\
Soient $i<n$, $\psi\force Yi$~; on doit montrer $S_U\star\psi\ps u_{n-i-1}\ps\pi\in\bbot$.
D'après la règle de réduction de~$S_U$, il suffit de montrer $\psi\star u_{n-i}\ps\pi\in\bbot$, ce qui
est évident, par hypothèse sur $\psi$.

\cqfd

\smallskip\noindent
{\bfseries Notation.} Pour chaque $n\in\NN$, on désigne par $\ul{n}$ l'entier de Church $\lbd f\lbd x(f)^nx$.\\
On pose $s=\lbd n\lbd f\lbd x(f)(n)fx$ (opération de successeur dans les entiers de Church).

\begin{theorem}[Mise en mémoire des entiers]\label{memoire_ent}
Soient $T,S\in\LLbd$ deux instructions avec les règles de réduction~: \ \
$T\star\phi\ps\nu\ps\pi\succ\nu\star S\ps\phi\ps \ul{0}[]\ps\pi$~; \
$S\star\psi\ps s^n\ul{0}[]\ps\pi\succ\psi\star s^{n+1}\ul{0}[]\ps\pi$.	 Alors~:\\
i) \ \ $T\force\pt X\pt n[(\{s^n\ul{0}[]\}\to X)\to($int$(n)\to X)]$\\
et $I[]\force\pt X\pt n[($int$(n)\to X)\to(\{s^n\ul{0}[]\}\to X)]$\\
lorsque $X$ est une variable propositionnelle.\\
ii) \ $T\force\pt X[\pt n(\{s^n\ul{0}[]\}\to Xn)\to\pt n($int$(n)\to Xn)]$\\
et $I[]\force\pt X[\pt n($int$(n)\to Xn)\to\pt n(\{s^n\ul{0}[]\}\to Xn)]$\\
lorsque $X$ est une variable de prédicat unaire.
\end{theorem}\noindent
Rappelons que la  notation $\xi\star\pi\succ\xi'\star\pi'$, utilisée pour les \gmg règles de réduction\gmd,
signifie simplement~: \  $\xi'\star\pi'\in\bbot$ \ $\Fl$ \ $\xi\star\pi\in\bbot$.

\smallskip\noindent
i) Pour la première formule, il suffit d'appliquer le théorème~\ref{memoire_gen} avec $u_n=s^n\ul{0}[]$.\\
Pour la seconde, on a \ $\vdash s^n\ul{0}:$ int$(n)$, donc $s^n\ul{0}[]\force$ int$(n)$ d'après le
théorème~\ref{adequat_gen} (lemme d'adéquation).
Soient alors $\xi\force$ int$(n)\to X$ et $\pi\in\|X\|$. On veut montrer~:\\
$\lbd x\,x[]\star\xi\ps s^n\ul{0}[]\ps\pi\in\bbot$. D'après la propriété~3 de $\bbot$, on doit montrer
$\xi'\star s^n\ul{0}[]\ps\pi\in\bbot$ pour tout $\xi'\le\xi$, ce qui est évident, puisque
$\xi'\star s^n\ul{0}[]\ps\pi\in\bbot$ par hypothèse sur $\xi$.\\
ii) Corollaire immédiat de (i).

\cqfd

\smallskip\noindent
La notation \ $\pt n\indi F[n]$ désigne la formule \ $\pt n($int$(n)\to F[n])$.
Le théorème~\ref{memoire_ent}(ii) permet de remplacer cette formule par \ $\pt n(\{s^n\ul{0}[]\}\to F[n])$.
Cette définition des \emph{quantificateurs res\-treints aux entiers} a l'avantage de permettre des calculs
de valeurs de formules beaucoup plus simples. C'est pourquoi nous l'utiliserons dans la suite.
Elle a néanmoins l'inconvénient de dépendre de la structure de réalisabilité considérée, par l'intermédiaire
de $\{s^n\ul{0}[]\}$, qui contient la substitution vide.

\smallskip\noindent
On considère un langage du second ordre pour l'arithmétique (avec symboles de fonctions, mais pas de constantes
de prédicat), et le modèle standard pour ce langage (modèle plein, les individus sont les entiers, chaque
symbole de fonction est interprété dans les entiers).\\
Alors toute équation vraie $\pt x_1\ldots\pt x_k(t(x_1,\ldots,x_k)=u(x_1,\ldots,x_k))$ ($t,u$ étant des termes
du langage) est réalisée par $I[{\cal S}]$. En restreignant les formules à \ int$(x)$, on remplace l'axiome de récurrence par les axiomes~:\\
\centerline{${\cal A}_f\equiv\pt x_1\ldots\pt x_k($int$(x_1),\ldots,\,$int$(x_k)\to$ int$(f(x_1,\ldots,x_k)))$}\\
pour chaque symbole de fonction $f$. On s'intéresse donc aux fonctions $f$ pour lesquelles ${\cal A}_f$ est réalisé (par une quasi-preuve close). Ce sont des fonctions récursives, puisque cette quasi-preuve calcule $f$.\\
Pour réaliser ${\cal A}_f$, il suffit de le démontrer, en logique du second ordre, à partir d'équations vraies
(qui comportent éventuellement des symboles de fonction autres que $f$). On peut ainsi utiliser les symboles pour
toutes les fonctions arithmétiques usuelles.

\smallskip\noindent
Pour le théorème~\ref{disj3} ci-dessous, on suppose l'existence des instructions suivantes~:\\
$\bullet$~~Etant donnés $\xi\in\LLbd$ et $\pi\in\PPi$, une instruction notée $\xi\kk_\pi$, avec la règle de réduction~:\\
\centerline{$\xi\kk_\pi\star\varpi\succ\xi\star\kk_\pi\ps\varpi$.}

\noindent
$\bullet$~~Une instruction notée $V$, avec la règle de réduction~:\\
\centerline{$V\star\xi\ps\eta\ps\pi\succ\eta\star\xi\kk_\pi\ps\pi$.}

\begin{theorem}\label{disj3}
Pour ${\cal X}\subset\PPi$, on définit \ ${\cal X}^-\subset\LLbd$ en posant \
${\cal X}^-=\{\kk_\pi;$ $\pi\in{\cal X}\}$.\\
Alors, si ${\cal X},{\cal Y}\subset\PPi$, on a~:\\
i) \ $I[{\cal S}]\force(\neg{\cal X}\to{\cal Y})\to({\cal X}^-\to{\cal Y})$~;\\
ii) \ $V\force({\cal X}^-\to{\cal Y})\to(({\cal Y}\to{\cal X})\to{\cal X})$.
\end{theorem}\noindent
\smallskip\noindent
i) Soient $\eta\force\neg{\cal X}\to{\cal Y}$, $\pi\in{\cal X}$ et $\rho\in{\cal Y}$.
On doit montrer que $I[\vsig]\star\eta\ps\kk_\pi\ps\rho\in\bbot$, pour toute substitution $\vsig\in{\cal S}$.
Or, on a $\vdash I:(\neg{\cal X\to{\cal Y}),\neg{\cal X}\to{\cal Y}}$ et donc~:\\
$I[\vsig]\force(\neg{\cal X\to{\cal Y}),\neg{\cal X}\to{\cal Y}}$ d'après le théorème~\ref{adequat_gen}
(lemme d'adéquation).\\
D'après le lemme~\ref{continuation}, on a $\kk_\pi\force\neg{\cal X}$, d'où le résultat.\\
ii) Soient $\xi\force{\cal X}^-\to{\cal Y}$,  $\eta\force{\cal Y}\to{\cal X}$ et $\pi\in{\cal X}$.
On doit montrer que \ $V\star\xi\ps\eta\ps\pi\in\bbot$, soit $\eta\star\xi\kk_\pi\ps\pi\in\bbot$.
Il suffit donc de montrer \ $\xi\kk_\pi\force{\cal Y}$~; soit $\varpi\in{\cal Y}$. On a
$\xi\star\kk_\pi\ps\varpi\in\bbot$, puisque $\kk_\pi\in {\cal X}^-$~; d'après la règle de réduction de
$\xi\kk_\pi$, on a donc \ $\xi\kk_\pi\star\varpi\in\bbot$.

\cqfd

\section*{Structures SR$_0$ et SR$_1$}
\smallskip\noindent
On considère un modèle de réalisabilité usuel (voir[3,4,5]), donné par un ensemble saturé $\bbot\subset\Lbd_c\fois\Pi$. Les éléments de $\Lbd_c$ sont les $\lbd$-termes clos comportant
les continuations $\kk_\pi$, les instructions $\ccc,\chi,\chi',\sigma$ (pour l'axiome du choix
non extensionnel, lemme~\ref{signature}), et éventuellement d'autres. Les instructions $T,S$
(pour la mise en mémoire des entiers, théorème~\ref{memoire_ent}) sont ici inutiles, car elles
sont données par les deux quasi-preuves~:

\smallskip
\centerline{$T_0=\lbd f\lbd n((n)\lbd g\lbd x(g)(s)x)f\ul{0}$ \ \ et \ $S_0=\lbd g\lbd x(g)(s)x$.}

\smallskip\noindent
Pour $\pi\in\Pi$ et $\tau\in\Lambda_c$, $\pi^\tau$ désigne la pile obtenue en ajoutant $\tau$
au fond de la pile $\pi$, juste avant la constante de pile~:\\
si $\pi=\xi_1\ps\ldots\ps\xi_n\ps\pi_0$, où $\pi_0$ est une constante de pile, alors $\pi^\tau=\xi_1\ps\cdots\ps\xi_n\ps\tau\ps\pi_0$.\\
Les nouvelles instructions $\chi,\chi'$ ont les règles de réduction suivantes~:\\
\centerline{$\chi\star\xi\ps\pi^\tau\succ\xi\star\tau\ps\pi$ \ (lecture)~; \
$\chi'\star\xi\ps\tau\ps\pi\succ\xi\star\pi^\tau$ \ (écriture).}

\smallskip\noindent
Cette structure de réalisabilité sera désignée par SR$_0$. On va en construire une nouvelle, que nous
désignerons par SR$_1$.

\smallskip\noindent
On considère un ensemble $P$ muni d'une opération binaire $(p,p')\mapsto pp'$, avec un
élément distingué $\1$. Pour chaque $p\in P$, on se donne un ensemble de termes $\C[p]\subset\Lambda_c$.\\
On suppose qu'on a cinq quasi-preuves closes $\alpha_i(0\le i\le4)$ telles que~:

\smallskip\noindent
$\tau\in\C[pq]$ $\Fl$ $\alpha_0\tau\in\C[p]$~; \
$\tau\in\C[pq]$ $\Fl$ $\alpha_1\tau\in\C[qp]$~;\\
$\tau\in\C[p]$ $\Fl$ $\alpha_2\tau\in\C[pp]$~; \
$\tau\in\C[p(qr)]$ $\Fl$ $\alpha_3\tau\in\C[(pq)r]$~;\\
$\tau\in\C[p]$ $\Fl$ $\alpha_4\tau\in\C[p\1]$.

\smallskip\noindent
Dans la suite, on considère des expressions de la forme $\gamma=(\alpha_{i_1})(\alpha_{i_2})\ldots(\alpha_{i_k})$\\
avec $0\le i_1,\ldots,i_k\le4$, qu'on appellera une \emph{composée de $\alpha_i$} $(0\le i\le4)$.\\
Une telle expression n'est pas dans $\Lbd_c$, mais $\gamma\tau\in\Lbd_c$ pour tout $\tau\in\Lbd_c$~;\\
le terme $\gamma\tau=(\alpha_{i_1})\ldots(\alpha_{i_k})\tau$ sera aussi écrit $(\gamma)\tau$.

\begin{lemma}\label{mono_comm_id}
On a six expressions $\gamma_0,\gamma_1,\gamma_2,\gamma_3,\gamma_4,\gamma_5$ qui sont des composées
de $\alpha_i$ ${(0\le i\le3)}$ telles que~:\\
$\tau\in\C[pq]$ $\Fl$ $\gamma_0\tau\in\C[p(pq)]$~; \
$\tau\in\C[(pq)r]$ $\Fl$ $\gamma_1\tau\in\C[pr]$~;\\
$\tau\in\C[(pq)r]$ $\Fl$ $\gamma_2\tau\in\C[qr]$~; \
$\tau\in\C[p(qr)]$ $\Fl$ $\gamma_3\tau\in\C[q(rr)]$~;\\
$\tau\in\C[p(qr)]$ $\Fl$ $\gamma_4\tau\in\C[qp]$~; \
$\tau\in\C[(pq)r]$ $\Fl$ $\gamma_5\tau\in\C[p(qr)]$.
\end{lemma}\noindent
On a $\tau\in\C[pq]$ $\Fl$ $(\alpha_2)\tau\in\C[(pq)(pq)]$ $\Fl$ $(\alpha_3)(\alpha_2)\tau\in\C[((pq)p)q)]$\\
$\Fl$ $(\alpha_0)(\alpha_3)(\alpha_2)\tau\in\C[(pq)p]$
$\Fl$ $(\alpha_1)(\alpha_0)(\alpha_3)(\alpha_2)\tau\in\C[p(pq)]$.\\
On pose donc $\gamma_0=(\alpha_1)(\alpha_0)(\alpha_3)(\alpha_2)$. On a~:\\
$\tau\in\C[(pq)r]$ $\Fl$ $(\alpha_1)\tau\in\C[r(pq)]$ $\Fl$ $(\alpha_3)(\alpha_1)\tau\in\C[(rp)q)]$
$\Fl$ $(\alpha_1)(\alpha_3)(\alpha_1)\tau\in\C[q(rp)]$\\
$\Fl$ $(\alpha_3)(\alpha_1)(\alpha_3)(\alpha_1)\tau\in\C[(qr)p)]$
$\Fl$ $(\alpha_1)(\alpha_3)(\alpha_1)(\alpha_3)(\alpha_1)\tau\in\C[p(qr))]$.\\
On pose donc $\gamma_5=(\alpha_1)(\alpha_3)(\alpha_1)(\alpha_3)(\alpha_1)$.\\
Calculs analogues pour les $\gamma_i$, $1\le i\le 4$.

\cqfd

\smallskip\noindent
{\small{\bfseries Remarque.} Soient $t(p_1,\ldots,p_n)$, $u(p_1,\ldots,p_n)$ deux termes écrits avec les
variables $p_1,\ldots,p_n$, un symbole de fonction binaire et un symbole de constante $\1$.\\
On montre facilement que, si la formule~: \
$\pt p_1\ldots\pt p_n[t(p_1,\ldots,p_n)\le u(p_1,\ldots,p_n)]$\\
est conséquence de~:\\
$\pt p\pt q\pt r((pq=qp)\land(p(qr)=(pq)r)\land(pp=p)\land(p\1=p))$, $\pt p\pt q(p\le q\dbfl pq=p)$,\\
alors, il existe une composée $\gamma$ des $\alpha_i(0\le i\le4)$ telle que l'on ait~:\\
$\tau\in\C[t(p_1,\ldots,p_n)]$ $\Fl$ $\gamma\tau\in\C[u(p_1,\ldots,p_n)]$.}

\smallskip\noindent
Pour chaque expression $\gamma$ qui est une composée de $\alpha_i$ $(0\le i\le4)$, on désigne par $\ov{\gamma}$ \ l'instruc\-tion dont la règle de réduction est~:\\
\centerline{$\ov{\gamma}\star\xi\ps\pi^\tau\succ\xi\star\pi^{\gamma\tau}$ pour $\xi\in\Lbd_c$ et $\pi\in\Pi$.}

\noindent
On peut, en fait, poser \ $\ov{\gamma}=\lbd x(\chi)\lbd y(\chi' x)(\gamma)y$.

\smallskip\noindent
On définit une nouvelle structure de réalisabilité en posant~:\\
$\LLbd=\Lambda_c\fois P$, $\PPi=\Pi\fois P$ et $\LLbd\star\PPi=(\Lbd_c\star\Pi)\fois P$.

\smallskip\noindent
Les opérations sont définies comme suit~:\\
$(\xi,p)\ps(\pi,q)=(\xi\ps\pi,pq)$~; \ $(\xi,p)\star(\pi,q)=(\xi\star\pi,pq)$~;\\
$\kk_{(\pi,p)}=(\kk^*_\pi,p)$ avec \ $\kk^*_\pi=\lbd x(\chi)\lbd y(\kk_\pi)(\chi' x)(\gamma_4)y$.

\smallskip\noindent
On prend pour ${\cal S}$ l'ensemble des substitutions de la forme $[(\xi_1,p)/x_1,\ldots,(\xi_k,p)/x_k]$.\\
Soit $t[x_1,\ldots,x_k]$ une quasi-preuve dont les variables libres sont parmi
$x_1,\ldots,x_k$. On pose~:\\
\centerline{$t[(\xi_1,p)/x_1,\ldots,(\xi_k,p)/x_k]=(t^*[\xi_1/x_1,\ldots,\xi_k/x_k],p)$}\\
où $t^*[x_1,\ldots,x_k]\in\QP$ est obtenu à partir de $t$ au moyen des règles suivantes~:\\
$x^*=x$~; \ $(tu)^*=\ov{\gamma}_0t^*u^*$~; \
$(\lbd x\,t)^*=\ov{\alpha}_3\lbd x\,t^*[\ov{\gamma}_2x/x,\ov{\gamma}_1x_1/x_1,\ldots,\ov{\gamma}_1x_k/x_k]$~;\\
$\ccc^*=\lbd x(\chi)\lbd y(\ccc)\lbd k((\chi' x)(\gamma_3)y)k^*$ \ avec \
$k^*=\lbd x(\chi)\lbd y(k)(\chi' x)(\gamma_4)y$.\\
Dans le cas de la substitution vide, notée $[]$, on a $t\in\QP_0$ et on pose $t[]=(t^*,\1)$.

\smallskip\noindent
Les $\lbd_c$-termes $\ccc^*$ et $\kk^*_\pi$ s'exécutent donc comme suit~:\\
\centerline{$\ccc^*\star\xi\ps\pi^\tau\succ\xi\star \kk^*_\pi\ps\pi^{\gamma_3\tau}$~; \
$\kk^*_\pi\star\xi\ps\varpi^\tau\succ\xi\star\pi^{\gamma_4\tau}$.}

\smallskip\noindent
On pose $\bbbot=\{(\xi\star\pi,p);$ $p\in P$, $(\pt\tau\in\C[p])\,\xi\star\pi^\tau\in\bbot\}$
($\bbot$ est le sous-ensemble saturé de $\Lbd_c\star\Pi$ qui définit la réalisabilité standard).

\smallskip\noindent
La structure de réalisabilité que nous venons de définir sera désignée par~SR$_1$.

\begin{lemma}\label{adequat_forcing}
$\bbbot$ est un ensemble saturé, c'est-à-dire qu'il a les propriétés~1 à~5, page~\pageref{compatibilite}.
\end{lemma}\noindent
1. Trivial, puisque $x_i[(\xi_1,p)/x_1,\ldots,(\xi_k,p)/x_k]=(\xi_i,p)$.\\
2. L'hypothèse est $t[(\xi'_1,p')/x_1,\ldots,(\xi'_k,p')/x_k]\star u[(\xi'_1,p')/x_1,\ldots,(\xi'_k,p')/x_k]\ps(\pi,q)\in\bbbot$ quels que soient 
$(\xi'_1,p')\le(\xi_1,p),\ldots,(\xi'_k,p')\le(\xi_k,p)$. En fait, on utilisera seulement le fait que
$t[(\xi_1,p)/x_1,\ldots,(\xi_k,p)/x_k]\star u[(\xi_1,p)/x_1,\ldots,(\xi_k,p)/x_k]\ps(\pi,q)\in\bbbot$\\
c'est-à-dire $(t^*[\xi_1/x_1,\ldots,\xi_k/x_k]\star u^*[\xi_1/x_1,\ldots,\xi_k/x_k]\ps\pi,p(pq))\in\bbbot$.\\
On doit montrer $tu[(\xi_1,p)/x_1,\ldots,(\xi_k,p)/x_k]\ps(\pi,q)\in\bbbot$, autrement dit~:\\
$((\ov{\gamma}_0t^*u^*)[\xi_1/x_1,\ldots,\xi_k/x_k]\star\pi,pq)\in\bbbot$.
On suppose donc $\tau\in\C[pq]$, donc $\gamma_0\tau\in\C[p(pq)]$ d'où
$t^*[\xi_1/x_1,\ldots,\xi_k/x_k]\star u^*[\xi_1/x_1,\ldots,\xi_k/x_k]\ps\pi^{\gamma_0\tau}\in\bbot$
et par suite~:\\
$(\ov{\gamma}_0t^*u^*)[\xi_1/x_1,\ldots,\xi_k/x_k]\star\pi^\tau\in\bbot$, ce qui est le résultat voulu.

\smallskip\noindent
3. On suppose $t[(\xi',p')/x,(\xi'_1,p')/x_1,\ldots,(\xi'_k,p')/x_k]\star(\pi,r)\in\bbbot$\\
quels que soient $(\xi',p')\le(\xi,q)$, $(\xi'_1,p')\le(\xi_1,p)$,\ldots,$(\xi_k',p')\le(\xi_k,p)$.\\
On doit montrer $\lbd x\,t[(\xi_1,p)/x_1,\ldots,(\xi_k,p)/x_k]\star(\xi,q)\ps(\pi,r)\in\bbbot$, c'est-à-dire~:\\
$((\lbd x\,t)^*[\xi_1/x_1,\ldots,\xi_k/x_k]\star\xi\ps\pi,p(qr))\in\bbbot$ ou encore~:\\
$(\ov{\alpha}_3\lbd x\,t^*[\ov{\gamma}_2x/x,\ov{\gamma}_1\xi_1/x_1,\ldots,\ov{\gamma}_1\xi_k/x_k]
\star\xi\ps\pi,p(qr))\in\bbbot$.\\
Soit donc $\tau\in\C[p(qr)]$. On doit montrer \
$\ov{\alpha}_3\lbd x\,t^*[\ov{\gamma}_2x/x,\ov{\gamma}_1\xi_1/x_1,\ldots,\ov{\gamma}_1\xi_k/x_k]\star
\xi\ps\pi^\tau\in\bbot$\\ soit \ $t^*[\ov{\gamma}_2\xi/x,\ov{\gamma}_1\xi_1/x_1,\ldots,\ov{\gamma}_1\xi_k/x_k]\star\pi^{\alpha_3\tau}\in\bbot$.

\begin{lemma}\label{force_le}
On a $(\ov{\gamma}_1\xi,pq)\le(\xi,p)$ et $(\ov{\gamma}_2\xi,pq)\le(\xi,q)$.
\end{lemma}\noindent
Supposons $(\xi,p)\star(\pi,r)\in\bbbot$, soit $(\xi\star\pi,pr)\in\bbbot$. On doit montrer
$(\ov{\gamma}_1\xi,pq)\star(\pi,r)\in\bbbot$, soit $(\ov{\gamma}_1\xi\star\pi,(pq)r)\in\bbbot$. On prend donc
$\tau\in\C[(pq)r]$, d'où $\gamma_1\tau\in\C[pr]$. On a donc $\xi\star\pi^{\gamma_1\tau}\in\bbot$, d'où
$\ov{\gamma}_1\xi\star\pi^\tau\in\bbot$, ce qui est le résultat voulu.\\
Même preuve pour la deuxième partie du lemme.

\cqfd

\smallskip\noindent
On a donc $(\ov{\gamma}_2\xi,pq)\le(\xi,q)$~; $(\ov{\gamma}_1\xi_i,pq)\le(\xi_i,p)$ pour $1\le i\le k$.
D'après l'hypothèse sur $t$, il en résulte que \ $t[(\ov{\gamma}_2\xi,pq)/x,(\ov{\gamma}_1\xi_1,pq)/x_1,\ldots,(\ov{\gamma}_1\xi_k,pq)/x_k]\star(\pi,r)\in\bbbot$,
ou encore~:\\
$(t^*[\ov{\gamma}_2\xi/x,\ov{\gamma}_1\xi_1/x_1,\ldots,\ov{\gamma}_1\xi_k/x_k]\star\pi,(pq)r)\in\bbbot$.\\
Or, on a \ $\tau\in\C[p(qr)]$, donc $\alpha_3\tau\in\C[(pq)r]$~; il en résulte que ~:\\
$t^*[\ov{\gamma}_2\xi/x,\ov{\gamma}_1\xi_1/x_1,\ldots,\ov{\gamma}_1\xi_k/x_k]\star\pi^{\alpha_3\tau}\in\bbot$
ce qui est le résultat voulu.

\smallskip\noindent
{\small{\bfseries Remarque.} Lorsque $k=0$ (cas de la substitution vide), la preuve reste valable, avec $p=\1$.}

\smallskip\noindent
4. On doit montrer~:\\
\centerline{$(\xi,q)\star\kk_{(\pi,r)}\ps(\pi,r)\in\bbbot$ $\Fl$
${\ccc[(\xi_1,p)/x_1,\ldots,(\xi_k,p)/x_k]\star(\xi,q)\ps(\pi,r)\in\bbbot}$}\\
ce qui s'écrit $(\xi,q)\star(\kk^*_\pi,r)\ps(\pi,r)\in\bbbot$ $\Fl$ 
$(\ccc^*,p)\star(\xi,q)\ps(\pi,r)\in\bbbot$, ou encore~:\\
$(\xi\star\kk^*_\pi\ps\pi,q(rr))\in\bbbot$ $\Fl$ $(\ccc^*\star\xi\ps\pi,p(qr))\in\bbbot$.\\
On suppose donc $\tau\in\C[p(qr)]$, d'où \ $\gamma_3\tau\in\C[q(rr)]$ et donc
$\xi\star\kk^*_\pi\ps\pi^{\gamma_3\tau}\in\bbot$.\\
D'après la règle de réduction pour $\ccc^*$, on a donc $\ccc^*\star\xi\ps\pi^\tau\in\bbot$, ce qui est
le résultat voulu.

\smallskip\noindent
5. $(\xi,q)\star(\pi,p)\in\bbbot$ $\Fl$ $(\kk^*_\pi,p)\star(\xi,q)\ps(\varpi,r)\in\bbbot$
c'est-à-dire~:\\
$(\xi\star\pi,qp)\in\bbbot$ $\Fl$ $(\kk^*_\pi\star\xi\ps\varpi,p(qr))\in\bbbot$.
On suppose donc $\tau\in\C[p(qr)]$, d'où $\gamma_4\tau\in\C[qp]$, d'où
$\xi\star\pi^{\gamma_4\tau}\in\bbot$. D'après la règle de réduction de $\kk^*_\pi$, on en déduit
$\kk^*_\pi\star\xi\ps\varpi^\tau\in\bbot$.

\cqfd

\smallskip\noindent
{\bfseries Notation.}\\
Dans SR$_1$, on notera $\vv F\vv$ la valeur de vérité d'une formule close $F$ (sous-ensemble de $\PPi=\Pi\fois P$) et \
$(\xi,p)\fforce F$ \ le fait que $(\xi,p)\in\LLbd$ réalise cette formule, c'est-à dire~:\\
$\pt\pi\pt q[(\pi,q)\in\vv F\vv$ $\Fl$ $(\xi\star\pi,pq)\in\bbbot]$.

\smallskip\noindent
Comme on veut utiliser, dans cette structure de réalisabilité, le théorème~\ref{memoire_ent} de mise en
mémoire des entiers, on doit trouver, dans $\LLbd$, les instructions nécessaires.\\
Elles s'écrivent \ $(T,\1),(S,\1)$, avec~:\\
$T=\ov{\beta}\lbd f\lbd n(n)Sf\ul{0}$~; $\beta$ est tel que \
$\tau\in\C[\1(p(qr))]$ $\Fl$ $\beta\tau\in\C[q(\1(p(\1r)))]$.\\
$S=\ov{\alpha}\lbd g\lbd x(g)(s)x$~; $\alpha$ est tel que \ $\tau\in\C[\1p]$ $\Fl$ $\alpha\tau\in\C[p]$.\\
On a alors, en effet, $s^n\ul{0}[]=(s^n\ul{0},\1)$ et~:\\
$(\nu,q)\star(S,\1)\ps(\phi,p)\ps(\ul{0},\1)\ps(\pi,r)\in\bbot$ \ $\Fl$ \
$(T,\1)\star(\phi,p)\ps(\nu,q)\ps(\pi,r)\in\bbot$.\\
$(\psi,p)\star(s^{n+1}\ul{0},\1)\ps(\pi,q)\in\bbot$ \ $\Fl$ \
$(S,\1)\star(\psi,p)\ps(s^n\ul{0},\1)\ps(\pi,q)\in\bbot$.

\smallskip\noindent
En effet, ces conditions s'écrivent~:\\
$(\nu\star S\ps\phi\ps\ul{0}\ps\pi,q(\1(p(\1r))))\in\bbot$ \ $\Fl$ \
$(T\star\phi\ps\nu\ps\pi,\1(p(qr)))\in\bbot$.\\
$(\psi\star s^{n+1}\ul{0}\ps\pi,p(\1q))\in\bbot$ \ $\Fl$ \
$(S\star\psi\ps s^n\ul{0}\ps\pi,\1(p(\1q)))\in\bbot$.

\smallskip\noindent
Noter que $\alpha,\beta$ sont des composées des $\alpha_i(0\le i\le3)$. L'élément $\1$ n'a aucune propriété
particulière et peut donc être pris quelconque dans $P$.

\smallskip\noindent
On veut également pouvoir utiliser le théorème~\ref{disj3}~; pour cela, on remarque que cette structure de
réalisabilité contient les instructions nécessaires. En effet~:

\smallskip\noindent
i) On définit une opération binaire sur $\LLbd$ en posant~:

\smallskip\noindent
\centerline{$(\xi,p)(\eta,q)=(\ov{\gamma}_5\xi\eta,pq)$ pour $(\xi,p),(\eta,q)\in\LLbd$.}\\
On a alors~:\\
\centerline{$(\xi,p)(\eta,q)\star(\pi,r)\succ(\xi,p)\star(\eta,q)\ps(\pi,r)$}
autrement dit~:\\
\centerline{$(\xi,p)\star(\eta,q)\ps(\pi,r)\in\bbbot$ $\Fl$ $(\xi,p)(\eta,q)\star(\pi,r)\in\bbbot$.}

\noindent
En effet, ceci s'écrit \ $(\xi\star\eta\ps\pi,p(qr))\in\bbbot$ $\Fl$
$(\ov{\gamma}_5\xi\eta\ps\pi,(pq)r)\in\bbbot$.\\
On suppose donc $\tau\in\C[(pq)r]$~; alors $\gamma_5\tau\in\C[p(qr)]$, donc
$\xi\star\eta\ps\pi^{\gamma_5\tau}\in\bbot$, d'où ${\ov{\gamma}_5\xi\eta\star\pi^\tau\in\bbot}$, ce qui est
le résultat voulu.

\smallskip\noindent
L'instruction cherchée est alors $(\xi,p)\kk_{(\pi,q)}$, soit $(\xi,p)(\kk^*_\pi,q)$ c'est-à-dire
$(\gamma_5\xi\kk^*_\pi,pq)$.

\smallskip\noindent
{\small{\bfseries Remarque.}
L'opération binaire que nous venons de définir dans $\LLbd$ \emph{n'est pas compatible} avec la substi\-tution
dans les quasi-preuves,  c'est-à-dire que~:\\
si $(\xi,p)=t[(\xi_1,p)/x_1,\ldots,(\xi_k,p)/x_k]$ et $(\eta,p)=u[(\xi_1,p)/x_1,\ldots,(\xi_k,p)/x_k]$,\\
alors $(\xi,p)(\eta,p)\ne tu[(\xi_1,p)/x_1,\ldots,(\xi_k,p)/x_k]$.}

\smallskip\noindent
ii) L'instruction cherchée sera notée $(V,\1)$ et doit avoir la règle de réduction suivante~:\\
\centerline{$(V,\1)\star(\xi,p)\ps(\eta,q)\ps(\pi,r)\succ(\eta,q)\star(\xi,p)\kk_{(\pi,r)}\ps(\pi,r)$}

\noindent
autrement dit \ \ $(\eta,q)\star(\xi,p)\kk_{(\pi,r)}\ps(\pi,r)\in\bbbot$ $\Fl$
$(V,\1)\star(\xi,p)\ps(\eta,q)\ps(\pi,r)\in\bbbot$.\\
Or, ceci s'écrit \ $(\eta\star\ov{\gamma}_5\xi\kk^*_\pi\ps\pi,q((pr)r))\in\bbbot$ $\Fl$
$(V\star\xi\ps\eta\ps\pi,\1(p(qr)))\in\bbbot$.\\
On suppose donc $\tau\in\C[\1(p(qr))]$ et on a donc $\gamma\tau\in\C[q((pr)r)]$ pour une composée
$\gamma$ convenable des $\alpha_i(0\le i\le3)$. On veut montrer $V\star\xi\ps\eta\ps\pi^\tau\in\bbot$
et on a ${\eta\star\ov{\gamma}_5\xi\kk^*_\pi\ps\pi^{\gamma\tau}\in\bbot}$.\\
On a donc \ $(\ov{\gamma}\eta)(\ov{\gamma}_5)\xi\kk^*_\pi\star\pi^\tau\in\bbot$ et on peut donc poser~:\\
\centerline{$V=(\chi)\lbd\tau\lbd x\lbd y(\ccc)\lbd k((\chi')(\ov{\gamma}y)(\ov{\gamma}_5)xk^*)\tau$}

\noindent
où $\gamma$ est une composée des $\alpha_i(0\le i\le3)$ telle que \
$\tau\in\C[\1(p(qr))]$ $\Fl$ $\gamma\tau\in\C[q((pr)r)]$\\
et \ $k^*=\lbd x(\chi)\lbd y(k)(\chi' x)(\gamma_4)y$.

\section*{Forcing}
Nous définissons ci-dessous deux classes de formules de formules du second ordre. Elles utilisent des
variables d'individu $x,y,\ldots$, des variables de prédicat $X,Y,\ldots,X^+,Y^+,\ldots$ \ (les variables
$X$ et $X^+$ ont la même arité), des variables de {\em condition} $p,q,\ldots$ \ Ce sont~:

\smallskip\noindent
1. Les \emph{formules de forcing}, ou \emph{formules de SR$_0$}, qui peuvent être interprétées dans SR$_0$ seulement.\\
2. Les \emph{formules de SR$_1$},  qui peuvent être interprétées dans SR$_1$ seulement.

\subsubsection*{1. Les formules de SR$_0$}
Les \emph{termes d'individu} sont construits avec les \emph{variables d'individu} $x_1,\ldots,x_k,\ldots$
et les symboles de fonction, qui sont toutes les fonction de $\NN^k$ dans $\NN$ ($k$ entier quelconque)
dont on aura besoin (récursives ou non). Un terme d'individu sera noté $t[x_1,\ldots,x_k]$ ou
$t\indi[x_1,\ldots,x_k]$ s'il y a ambiguïté.\\
Les \emph{termes de condition} sont construits avec les \emph{variables de condition} $p_1,\ldots,p_k,\ldots$,
la cons\-tante $\1$ et un symbole de fonction binaire (par exemple $((pq)\1)(qp)$ \ est un terme de condition).
Un terme de condition sera noté $t[p_1,\ldots,p_k]$ ou $t^{\PP}[p_1,\ldots,p_k]$ s'il y a ambiguïté.

\smallskip\noindent
Les \emph{formules de SR$_0$} sont construites comme suit~:\\
$\bullet$~~Les formules de SR$_0$ {\em atomiques}, de la forme~:\\
$t\indi\ne u\indi$, $X(t\indi_1,\ldots,t\indi_n)$, $t^{\PP}\ne u^{\PP}$ et
$t^{\PP}\neps X^+(t\indi_1,\ldots,t\indi_n)$, où $X$ est
une variable de prédicat d'arité $n$, $t\indi,u\indi,t\indi_1,\ldots,t\indi_n$ sont des termes
d'individu et $t^{\PP},u^{\PP}$ des termes de condition.\\
La formule atomique $t^{\PP}\neps X^+(t\indi_1,\ldots,t\indi_n)$ \ se lit \gmg la condition
$t^{\PP}$ \emph{s'oppose à} $X^+(t\indi_1,\ldots,t\indi_n)$\gmd.\\
$\bullet$~~Si $F$ et $G$ sont des formules de SR$_0$, alors $F\to G$ en est une aussi.\\
$\bullet$~~Si $F$ est une formule de SR$_0$, alors $\C[t^{\PP}]\to F$ en est une aussi.\\
$\bullet$~~Si $F$ est une formule de SR$_0$, alors $\pt x\,F$, $\pt x\indi F$ et $\pt p^{\PP}\,F$ sont des formules
de SR$_0$ ($x$~est une variable d'individu, $p$ une variable de condition).\\
$\bullet$~~Si $F$ est une formule de SR$_0$ et $X$ une variable de prédicat, alors $\pt X\,F$ et $\pt X^+\,F$ sont des formules de SR$_0$.\\
{\small{\bfseries Remarque.} $\top$ et $\bot$ se trouvent parmi les formules atomiques~: ce sont
$0\ne1$ et $0\ne0$.}

\smallskip\noindent
On utilisera aussi deux sous-classes~:\\
i) Les formules de SR$_0$ {\em usuelles}, construites avec les mêmes règles, sauf que l'on
ne permet que les formules atomiques $t\indi\ne u\indi$, $X(t\indi_1,\ldots,t\indi_n)$ et les
quantificateurs $\pt x$, $\pt x\indi$ et $\pt X$.\\
ii) Les formules de SR$_0$ {\em usuelles restreintes à $\NN$}, sous-classe de (i) où l'on ne permet que
les quantificateurs $\pt x\indi$ et $\pt X$.

\smallskip\noindent
Les formules de SR$_0$ prennent une valeur de vérité dans SR$_0$, c'est-à-dire dans ${\cal P}(\Pi)$~;
les va\-riables de prédicat $X$ va\-rient dans SR$_0$, les variables $X^+$ dans SR$_1$.\\
Soit donc $F$ une formule de SR$_0$ \emph{close avec paramètres}~: il s'agit d'une formule de SR$_0$,
dans laquelle les variables libres d'individu ont été substituées par des entiers, les variables libres
de condition ont été substituées par des éléments de $P$~; les variables libres $X$ de prédicat $n$-aire
ont été substituées par des prédicats de SR$_0$, c'est-à-dire des éléments de ${\cal P}(\Pi)^{\NN^n}$~;
les variables libres $X^+$ de prédicat $n$-aire ont été substituées par des
prédicats de SR$_1$, c'est-à-dire des éléments de ${\cal P}(\Pi\fois P)^{\NN^n}$.\\
On définit sa valeur de vérité $\|F\|\subset\Pi$ par récurrence sur la construction de $F$ (comme d'habitude,
$\xi\force F$ signifie $(\pt\pi\in\|F\|)\,\xi\star\pi\in\bbot$)~:

\smallskip\noindent
$\bullet$~~Si $F$ est atomique de la forme $t\ne u$, alors $t,u$ sont des termes clos d'individu ou de condition.
Ils ont donc une valeur dans $\NN$ ou $P$~; $\|t\ne u\|$ est $\vide$ ou $\Pi$ suivant que ces valeurs sont
distinctes ou non.\\
$\bullet$~~Si $F$ est atomique de la forme $X(t_1,\ldots,t_n)$, alors $X\in{\cal P}(\Pi)^{\NN^n}$ et
$t_1,\ldots,t_n$ sont des termes d'individu clos, donc ont une valeur dans $\NN\;$. $\|X(t_1,\ldots,t_n\|$ est
alors défini de façon évidente.\\
$\bullet$~~Si $F$ est atomique de la forme $t^{\PP}\neps X^+(t_1,\ldots,t_n)$, alors
$X^+\in{\cal P}(\Pi\fois P)^{\NN^n}$ et $t_1,\ldots,t_n$ sont des termes d'individu clos, donc ont une valeur
dans $\NN\;$. On a donc ${X^+(t_1,\ldots,t_n)\subset\Pi\fois P}$.\\
Or, $t^{\PP}$ est un terme de condition clos, donc prend une valeur $q\in P$. On pose alors~:\\
$\|t^{\PP}\neps X^+(t_1,\ldots,t_n)\|=\|F\|=\{\pi\in\Pi;$ $(\pi,q)\in X^+(t_1,\ldots,t_n)\}$.\\
$\bullet$~~Si $F\equiv G\to H$, alors \ $\|F\|=\{\xi\ps\pi;$ $\xi\force G,\pi\in\|H\|\}$.\\
$\bullet$~~Si $F\equiv\C[t^{\PP}]\to H$, alors \ $\|F\|=\{\xi\ps\pi;$ $\xi\in\C[q],\pi\in\|H\|\}$
($q\in P$ est la valeur de $t^{\PP}$).\\ 
$\bullet$~~Si $F\equiv\pt x\,G$, alors $\dsp\|F\|=\bigcup_{n\in\NN}\|G[n/x]\|$.\\
$\bullet$~~Si $F\equiv\pt x\indi G$, alors $\dsp\|F\|=\bigcup_{n\in\NN}\{s^n\ul{0}\ps\pi;$ $\pi\in\|G[n/x]\|\}$.\\
$\bullet$~~Si $F\equiv\pt p^P\,G$, alors $\dsp\|F\|=\bigcup_{p\in P}\|G[p]\|$.\\
$\bullet$~~Si $F\equiv\pt X\,G$, où $X$ est une variable de prédicat $n$-aire, alors
$\dsp\|F\|=\hspace{-1em}\bigcup_{\ov{X}\in{\cal P}(\Pi)^{\NN^n}}\hspace{-1em}\|G[\ov{X}/X]\|$.\\
$\bullet$~~Si $F\equiv\pt X^+G$, où $X$ est une variable de prédicat $n$-aire, alors
$\dsp\|F\|=\hspace{-1em}\bigcup_{{\cal X}\in{\cal P}(\Pi\fois P)^{\NN^n}}\hspace{-1em}\|G[{\cal X}/X]\|$.

\smallskip\noindent
Le lemme~\ref{leibniz} montre que la formule $t\ne u$ a bien la signification voulue.
\begin{lemma}\label{leibniz}\ \\
Soit $t=u$ la formule $t\ne u\to\bot$. Alors, pour toute formule $F$ de SR$_0$, on a~:\\
$\lbd x\lbd y(\ccc)\lbd k(x)(k)y\force\pt x\indi\pt y\indi[x=y,F(x)\to F(y)]$~;\\
$\lbd x\lbd y(\ccc)\lbd k(x)(k)y\force\pt p^{\PP}\pt q^{\PP}[p=q,F(p)\to F(q)]$.
\end{lemma}\noindent
Les deux énoncé se démontrent de la même façon. On montre le deuxième~:\\
Soient $p,q\in P$~; $\xi\force p\ne q\to\bot$~; $\eta\force F(p)$ et $\pi\in\|F(q)\|$. On doit montrer que~:\\
$\lbd x\lbd y(\ccc)\lbd k(x)(k)y\star\xi\ps\eta\ps\pi\in\bbot$, soit $\xi\star\kk_\pi\eta\ps\pi\in\bbot$.\\
C'est immédiat si $p\ne q$, car alors $\xi\force\top\to\bot$. Si $p=q$, alors $\|F(p)\|=\|F(q)\|$, donc
$\eta\star\pi\in\bbot$, d'où $\kk_\pi\eta\force\bot$. Mais $\xi\force\bot\to\bot$, d'où le résultat.

\cqfd

\subsubsection*{2. Les formules de SR$_1$}
Les \emph{formules de SR$_1$} sont construites comme suit~:\\
$\bullet$~~Les formules de SR$_1$ {\em atomiques}, de la forme \ $t\ne u$, $X(t_1,\ldots,t_n)$ et
$X^+(t_1,\ldots,t_n)$, où $X$ est une variable de prédicat d'arité $n$ et $t,u,t_1,\ldots,t_n$ sont des termes d'individu.\\
$\bullet$~~Si $F$ et $G$ sont des formules de SR$_1$, alors $F\to G$ en est une aussi.\\
$\bullet$~~Si $F$ est une formule de SR$_1$, alors $\pt x\,F$ et $\pt x\indi F$ sont des formules
de SR$_1$.\\
$\bullet$~~Si $F$ est une formule de SR$_1$ et $X$ une variable de prédicat, alors $\pt X\,F$ et $\pt X^+\,F$ sont des formules de SR$_1$.

\smallskip\noindent
{\small{\bfseries Remarque.} Les formules usuelles de SR$_0$ sont dans cette classe~: ce sont celles qui ne
comportent pas de formule atomique de la forme $X^+(t_1,\ldots,t_n)$ ni de quantificateur $\pt X^+$.}

\smallskip\noindent
On utilisera aussi une sous-classe~:\\
Les formules de SR$_1$ {\em restreintes à $\NN$}, construites avec les mêmes règles, sauf que l'on
ne permet que les quantificateurs $\pt x\indi$, $\pt X$ et $\pt X^+$.\\
{\small{\bfseries Remarque.} Les formules restreintes de SR$_0$ sont dans cette classe~: ce sont celles qui ne
comportent pas de formule atomique de la forme $X^+(t_1,\ldots,t_n)$ ni de quantificateur $\pt X^+$.}

\smallskip\noindent
Les formules de SR$_1$ prennent une valeur de vérité dans SR$_1$, c'est-à-dire dans
${\cal P}(\Pi\fois P)$~; les variables de prédicat $X$ va\-rient dans SR$_0$, les variables
$X^+$ dans SR$_1$.\\
Soit donc $F$ une formule de SR$_1$ \emph{close avec paramètres}~: il s'agit d'une formule de SR$_1$,
dans laquelle les variables libres d'individu ont été substituées par des entiers, les variables
libres~$X$ de prédicat $n$-aire ont été substituées par des prédicats de SR$_0$, c'est-à-dire des éléments de
${\cal P}(\Pi)^{\NN^n}$~; les variables libres $X^+$ de prédicat $n$-aire ont été substituées par des
prédicats de SR$_1$, c'est-à-dire des éléments de ${\cal P}(\Pi\fois P)^{\NN^n}$.\\
On définit sa valeur de vérité $\vv F\vv\subset\Pi\fois P$ par récurrence sur la construction de $F$ (bien
entendu, $(\xi,p)\fforce F$ \ signifie \ $\pt\pi\pt q[(\pi,q)\in\vv F\vv$ $\Fl$ $(\xi,p)\star(\pi,q)\in\bbbot]$)~:

\smallskip\noindent
$\bullet$~~Si $F$ est atomique de la forme $t\ne u$, alors $t,u$ sont des termes d'individu clos, donc ont
une valeur dans $\NN$~; $\vv t\ne u\vv$ est $\vide$ ou $\Pi\fois P$ suivant que ces entiers sont distincts ou non.\\
$\bullet$~~Si $F$ est atomique de la forme $X^+(t_1,\ldots,t_n)$, alors
$X^+\in{\cal P}(\Pi\fois P)^{\NN^n}$ et $t_1,\ldots,t_n$ sont des termes clos, donc ont une valeur dans $\NN$.
On a donc $X^+(t_1,\ldots,t_n)\subset\Pi\fois P$~; cela définit $\vv X^+(t_1,\ldots,t_n)\vv$.\\
$\bullet$~~Si $F$ est atomique de la forme $X(t_1,\ldots,t_n)$, alors $X\in{\cal P}(\Pi)^{\NN^n}$ et
$t_1,\ldots,t_n$ sont des termes clos, donc ont une valeur dans $\NN$. La valeur $\|X(t_1,\ldots,t_n)\|$
{\em dans SR$_0$} est alors définie de façon évidente. Sa valeur $\vv X(t_1,\ldots,t_n)\vv$ {\em dans SR$_1$}
est définie par \ $\vv X(t_1,\ldots,t_n)\vv=\|X(t_1,\ldots,t_n)\|\fois P$.\\
$\bullet$~~Si $F\equiv G\to H$, alors \ $\vv F\vv=\{(\xi,p)\ps(\pi,q);$ $(\xi,p)\fforce G,\,
(\pi,q)\in\vv H\vv\}$.\\
$\bullet$~~Si $F\equiv\pt x\,G$, alors $\dsp\vv F\vv=\bigcup_{n\in\NN}\vv G[n/x]\vv$.\\
$\bullet$~~Si $F\equiv\pt x\indi G$, alors $\dsp\vv F\vv=\bigcup_{n\in\NN}\{(s^n\ul{0},\1)\ps(\pi,q);$
$(\pi,q)\in\vv G[n/x]\vv\}$.\\
$\bullet$~~Si $F\equiv\pt X^+G$, où $X$ est une variable de prédicat $n$-aire, alors
$\dsp\vv F\vv=\hspace{-1em}\bigcup_{{\cal X}\in{\cal P}(\Pi\fois P)^{\NN^n}}\hspace{-1em}\vv G[{\cal X}/X]\vv$.

\smallskip\noindent
$\bullet$~~Si $F\equiv\pt X\,G$, où $X$ est une variable de prédicat $n$-aire, alors
$\dsp\vv F\vv=\hspace{-1em}\bigcup_{\ov{X}\in{\cal P}(\Pi)^{\NN^n}}\hspace{-1em}\vv G[\ov{X}/X]\vv$.
 
\subsubsection*{Définition du forcing}
Considérons une formule $F$ de SR$_1$ et soit $p$ une variable de condition.\\
On définit, par récurrence sur $F$, une formule de forcing (formule de SR$_0$) notée
$p\force^fF$ (lire \gmg$p$ force $F$\gmd).

\smallskip\noindent
$\bullet$~~Si $F$ est atomique de la forme $X^+(t_1,\ldots,t_n)$, on pose~:\\
\centerline{$p\force^f X(t_1,\ldots,t_n)\;\equiv\;\pt q^P\{\C[pq]\to q\neps X^+(t_1,\ldots,t_n)\}$.}

\smallskip\noindent
$\bullet$~~Si $F$ est atomique de la forme $X(t_1,\ldots,t_n)$, on pose~:\\
\centerline{$p\force^f X(t_1,\ldots,t_n)\;\equiv\;\C[p]\to X(t_1,\ldots,t_n)$.}

\smallskip\noindent
$\bullet$~~Si $F$ est atomique de la forme $t\ne u$, on pose~:\\
\centerline{$p\force^f t\ne u\;\equiv\;\C[p]\to t\ne u$.}

\smallskip\noindent
$\bullet$~~$p\force^fG\to H\;\equiv\;\pt q^P\{(q\force^fG)\to(pq\force^fH)\}$.\\
$\bullet$~~$p\force^f\pt x\,G\;\equiv\;\pt x(p\force^fG)$.\\
$\bullet$~~$p\force^f\pt x\indi G\;\equiv\;\pt x\indi(p\force^fG)$.\\
$\bullet$~~$p\force^f\pt X^+\,G\;\equiv\;\pt X^+(p\force^fG)$.\\
$\bullet$~~$p\force^f\pt X\,G\;\equiv\;\pt X(p\force^fG)$.

\smallskip\noindent
On considère, dans la suite, des formules de SR$_1$ closes, à paramètres ${\cal X}_1,\ldots,{\cal X}_k$
dans SR$_1$ et $\ov{Y}_1,\ldots,\ov{Y}_l$ dans SR$_0$.\\
On les écrira \ $F[{\cal X}_1,\ldots,{\cal X}_k,\ov{Y}_1,\ldots,\ov{Y}_l]$, au lieu de
$F[{\cal X}_1/X^+_1,\ldots,{\cal X}_k/X^+_k,\ov{Y}_1/Y_1,\ldots,\ov{Y}_l/Y_l]$.\\
Un paramètre (constante de prédicat d'arité $n$) de SR$_1$ est une application
${\cal X}:\NN^n\to{\cal P}(\Pi\fois P)$.\\
Un paramètre de SR$_0$ est une application $\ov{Y}:\NN^n\to{\cal P}(\Pi)$.\\
On écrira donc, en abrégé, \ $p\force^fF[{\cal X}_1,\ldots,{\cal X}_k,\ov{Y}_1,\ldots,\ov{Y}_l]$.\\
Par exemple, si ${\cal X}\in{\cal P}(\Pi\fois P)^{\NN}$ (prédicat unaire de SR$_1$), alors~:\\
$p\force^f{\cal X}n$ est la formule de forcing \ $\pt q[\C[pq]\to q\neps{\cal X}n]$.

\smallskip\noindent
{\small{\bfseries Remarque.} Une formule atomique de la forme $t\ne u$ est alors de la forme $\ov{X}(t,u)$
où $\ov{X}:\NN^2\to{\cal P}(\Pi)$ est défini par $\ov{X}(m,n)=\vide$ si $m\ne n$ et
$\ov{X}(m,n)=\Pi$ si $m=n$.}

\smallskip\noindent
Etant donnés une telle formule $F[{\cal X}_1,\ldots,{\cal X}_k,\ov{Y}_1,\ldots,\ov{Y}_l]$ et $p\in P$,
on définit une valeur de vérité de SR$_0$ (sous-ensemble de $\Pi$), notée $\|p\force'F\|$.\\
On pose~:\\
\centerline{$\dsp\|p\force'F\|=\bigcup_{q\in P}\{\pi^\tau;$ $\tau\in\C[pq],(\pi,q)\in\vv F\vv\}$.}

\begin{lemma}\label{fforce'}
Pour $\xi\in\Lbd_c$ et $p\in P$, on a \ $(\xi,p)\fforce F$ \ $\Dbfl$ \ $\xi\force(p\force'F)$.
\end{lemma}\noindent
En effet, la condition \ $(\xi,p)\fforce F$ \ s'écrit \
$\pt\pi\pt q\{(\pi,q)\in\vv F\vv$ $\Fl$ $(\xi,p)\star(\pi,q)\in\bbbot\}$, c'est-à-dire
$\pt\pi\pt q\pt\tau\{q\in P\,,\tau\in\C[pq],(\pi,q)\in\vv F\vv$ $\Fl$ $\xi\star\pi^\tau\in\bbot\}$.
Par définition de $\|p\force'F\|$, cette condition équivaut à \
$\pt\varpi(\varpi\in\|p\force'F\|$ $\Fl$ $\xi\star\varpi\in\bbot)$.

\cqfd

\begin{theorem}\ \\
Pour chaque formule close $F[{\cal X}_1,\ldots,{\cal X}_k,\ov{Y}_1,\ldots,\ov{Y}_l]$ de SR$_1$,
à paramètres ${\cal X}_1,\ldots,{\cal X}_k$ dans SR$_1$ et $\ov{Y}_1,\ldots,\ov{Y}_l$ dans SR$_0$,
il existe deux quasi-preuves closes $\chi_F,\chi'_F$ telles que~:\\
\centerline{$\chi_F\force[(p\force^fF)\to(p\force'F)]$ \ et \ $\chi'_F\force[(p\force'F)\to(p\force^fF)]$.}

\noindent
$\chi_F$ et $\chi'_F$ ne dépendent pas des paramètres et des termes présents dans~$F$.
\end{theorem}\noindent
On le montre par récurrence sur la construction de la formule $F$.

\smallskip\noindent
$\bullet$~~Cas où $F\equiv\pt x\,G$. On a alors~:\\
$\dsp\|p\force'F\|=\bigcup_{q\in P}\{\pi^\tau;$ $\dsp\tau\in\C[pq],(\pi,q)\in\vv\pt x\,G\vv\}=
\bigcup_{q\in P,n\in\NN}\{\pi^\tau;$ $\tau\in\C[pq],(\pi,q)\in\vv G[n/x]\vv\}$,\\
et donc~:\\
\centerline{$\dsp\|p\force'F\|=\bigcup_{n\in\NN}\|p\force'G[n/x]\|$.}

\noindent
On a aussi~:\\
\centerline{$\dsp\|p\force^fF\|=\bigcup_{n\in\NN}\|p\force^fG[n/x]\|$.}

\noindent
L'hypothèse de récurrence donne alors immédiatement le résultat, si on pose~:\\
\centerline{$\chi_F=\chi_G$ et $\chi'_F=\chi'_G$.}

\smallskip\noindent
$\bullet$~Cas où $F\equiv\pt X\,G$ ou bien $F\equiv\pt X^+\,G$. Même démonstration. On a encore~:\\
\centerline{$\chi_F=\chi_G$ et $\chi'_F=\chi'_G$.}

\smallskip\noindent
$\bullet$~~Cas où $F\equiv\pt x\indi G$.\\
On a $\dsp\|p\force'F\|=\bigcup_{q\in P}\{\varpi^\tau;$
$\dsp\tau\in\C[pq],(\varpi,q)\in\vv\pt x\indi G\vv\}=\\
\bigcup_{q\in P,n\in\NN}\{\varpi^\tau;$
$\dsp\tau\in\C[pq],(\varpi,q)\in\vv\{(s^n\ul{0},\1)\}\to G[n]\vv\}$. On a donc~:\\
$(1)$\hspace{7em}$\dsp\|p\force'F\|=\bigcup_{q\in P,n\in\NN}\{s^n\ul{0}\ps\pi^\tau;$
$\tau\in\C[p(\1q)],(\pi,q)\in\vv G[n]\vv\}$.

\noindent
Par ailleurs, par définition de $\|p\force^fF\|$, on a~:

\smallskip\noindent
$(2)$\hspace{4em}$\dsp\|p\force^fF\|=\bigcup_{n\in\NN}\|\{s^n\ul{0}\}\to(p\force^fG[n])\|=
\bigcup_{n\in\NN}\{s^n\ul{0}\ps\pi;$ $\pi\in\|p\force^fG[n]\|\}$.

\smallskip\noindent
i) Par hypothèse de récurrence, on a \ $\chi_G\force(p\force^fG[n])\to(p\force'G[n])$. Posons~:\\
\centerline{$\chi_F=\lbd x\lbd n(\ov{\alpha}\chi_G)(x)n$}

\noindent
où $\alpha$ est une composée des
$\alpha_i(0\le i\le3)$ telle que \ $\tau\in\C[p(\1q)]$ $\Fl$ $\alpha\tau\in\C[pq]$.\\
Soient $\xi\force(p\force^fF)$ et $s^n\ul{0}\ps\pi^\tau\in\|p\force'F\|$,
avec $\tau\in\C[p(\1q)]$ et $(\pi,q)\in\vv G[n]\vv$ (d'après~$(1)$). On doit montrer
$\chi_F\star\xi\ps s^n\ul{0}\ps\pi^\tau\in\bbot$.\\
Or, on a $\alpha\tau\in\C[pq]$, donc $\pi^{\alpha\tau}\in\|p\force'G[n]\|$, par définition de $\|p\force'G[n]\|$.\\
On a aussi $\xi\force(\{s^n\ul{0}\}\to(p\force^f G[n]))$, d'après $(2)$ et l'hypothèse sur $\xi$.\\
Donc \ $(\xi)s^n\ul{0}\force(p\force^fG[n])$~; l'hypothèse de récurrence donne alors~:\\
$\chi_G\star(\xi)s^n\ul{0}\ps\pi^{\alpha\tau}\in\bbot$. On a donc bien
$\chi_F\star\xi\ps s^n\ul{0}\ps\pi^\tau\in\bbot$, d'après le choix de $\chi_F$.

\smallskip\noindent
ii) Par hypothèse de récurrence, on a \ $\chi'_G\force(p\force'G[n])\to(p\force^fG[n])$. Posons~:\\
\centerline{$\chi'_F=\lbd x\lbd n(\chi'_G)(\ov{\alpha})xn$}

\noindent
où $\alpha$ est une composée des $\alpha_i(0\le i\le4)$ telle que \
$\tau\in\C[pq]$ $\Fl$ $\alpha\tau\in\C[p(\1q)]$.\\
Soient $\xi\force(p\force'F)$ et $s^n\ul{0}\ps\pi\in\|p\force^fF\|$, avec $\pi\in\|p\force^fG[n]\|$
(d'après $(2)$).\\
On doit montrer $\chi'_F\star\xi\ps s^n\ul{0}\ps\pi\in\bbot$.\\
Or, d'après $(1)$, on a $\xi\star s^n\ul{0}\ps\pi^\tau\in\bbot$ si $(\pi,q)\in\vv G[n]\vv$ et
$\tau\in\C[p(\1q)]$. On a donc~:\\
$(\ov{\alpha}\xi)s^n\ul{0}\star\pi^\tau\in\bbot$ si $(\pi,q)\in\vv G[n]\vv$ et $\tau\in\C[pq)]$ (car alors
$\alpha\tau\in\C[p(\1q)]$).\\
Il en résulte que \ $(\ov{\alpha}\xi)s^n\ul{0}\force(p\force'G[n])$. L'hypothèse de récurrence donne alors~:\\
$\chi'_G\star(\ov{\alpha}\xi)s^n\ul{0}\ps\pi\in\bbot$, d'où le résultat.

\smallskip\noindent
$\bullet$~~Cas où $F\equiv G\to H$.\\
i) Soit $\alpha$ une composée des $\alpha_i(0\le i\le3)$ telle que \
$\tau\in\C[p(qr)]$ $\Fl$ $\alpha\tau\in\C[(pq)r]$.\\
Soient $\xi\force(p\force^fG\to H)$ et $\varpi^\tau\in\|p\force'(G\to H)\|$.\\
On a donc \ $\tau\in\C(pq')$ et $(\varpi,q')\in\vv G\to H\vv$.\\
Donc $\varpi=\eta\ps\pi$, $q'=qr$, $(\eta,q)\fforce G$ et $(\pi,r)\in\vv H\vv$.\\
D'après le lemme~\ref{fforce'}, on a \ $\eta\force(q\force'G)$. L'hypothèse de récurrence
donne donc~:\\
$\chi'_G\eta\force(q\force^fG)$, d'où $(\xi)(\chi'_G)\eta\force (pq\force^fH)$. L'hypothèse de récurrence
donne alors~:\\
$(\chi_H)(\xi)(\chi'_G)\eta\force(pq\force'H)$, donc par le lemme~\ref{fforce'},
$((\chi_H)(\xi)(\chi'_G)\eta,pq)\fforce H$.\\
Il en résulte que $((\chi_H)(\xi)(\chi'_G)\eta\star\pi,(pq)r)\in\bbbot$.\\
Comme $\tau\in\C[p(qr)]$, on a \ $\alpha\tau\in\C[(pq)r)]$ et donc
$(\chi_H)(\xi)(\chi'_G)\eta\star\pi^{\alpha\tau}\in\bbot$.\\
Posons~:\\
\centerline{$\chi_F=(\ov{\alpha})\lbd x\lbd y(\chi_H)(x)(\chi'_G)y$.}

\smallskip\noindent
On a alors \ $\chi_F\star\xi\ps\varpi^\tau=\chi_F\star\xi\ps\eta\ps\pi^\tau\succ
(\chi_H)(\xi)(\chi'_G)\eta\star\pi^{\alpha\tau}\in\bbot$ \ et donc~:\\
$\chi_F\force[(p\force^fG\to H)\to(p\force'G\to H)]$.

\smallskip\noindent
ii) A l'aide de l'hypothèse de récurrence, on montre d'abord le:

\begin{lemma}\label{alpha_xi_chi_G_eta}
Si $\xi\force(p\force'G\to H)$ et $\eta\force(q\force^fG)$ alors
$(\ov{\alpha}\xi)(\chi_G)\eta\force(pq\force'H)$~;\\
$\alpha$ est une composée des $\alpha_i(0\le i\le3)$ telle que \
$\tau\in\C[(pq)r]$ $\Fl$ $\alpha\tau\in\C[p(qr)]$.
\end{lemma}\noindent
Soit $\rho^\tau\in\|pq\force'H\|$, c'est-à-dire qu'on a $\tau\in\C[(pq)r]$ et $(\rho,r)\in\vv H\vv$.\\
Par hypothèse de récurrence, on a $\chi_G\eta\force(q\force'G)$ et donc $(\chi_G\eta,q)\fforce G$
d'après le lemme~\ref{fforce'}.\\
On a donc $(\chi_G\eta\ps\rho,qr)\in\vv G\to H\vv$.\\
Par ailleurs, on a $\alpha\tau\in\C[p(qr)]$ et donc $\chi_G\eta\ps\rho^{\alpha\tau}\in\|p\force'G\to H\|$.\\
Il en résulte que \ $\xi\star\chi_G\eta\ps\rho^{\alpha\tau}\in\bbot$ et donc \
$(\ov{\alpha}\xi)(\chi_G)\eta\star\rho^\tau\in\bbot$.

\cqfd

\smallskip\noindent
Soient alors $\xi\force(p\force'G\to H)$, $\eta\force(q\force^fG)$ et $\pi\in\|pq\force^fH\|$.\\
Par hypothèse de récurrence, on a \ $\chi'_H\force(pq\force'H)\to(pq\force^fH)$.\\
D'après le lemme~\ref{alpha_xi_chi_G_eta}, il en résulte que \
$\chi'_H\star(\ov{\alpha}\xi)(\chi_G)\eta\ps\pi\in\bbot$~; on a donc \
$\chi'_F\star\xi\ps\eta\ps\pi\in\bbot$ avec \ $\chi'_F=\lbd x\lbd y(\chi'_H)(\ov{\alpha}x)(\chi_G)y$.
 
\smallskip\noindent
$\bullet$~~Cas où $F\equiv{\cal X}(t_1,\ldots,t_k)$. On a alors~:\\
$\dsp\|p\force^fF\|=\bigcup_{q\in P}\{\tau\ps\pi;$ $\tau\in\C[pq]~,(\pi,q)\in\vv{\cal X}(t_1,\ldots,t_k)\vv\}$ et\\
$\dsp\|p\force'F\|=\bigcup_{q\in P}\{\pi^\tau;$ $\tau\in\C[pq]~,(\pi,q)\in\vv{\cal X}(t_1,\ldots,t_k)\vv\}$.\\
On a donc \ $\chi_F=\chi$ \ et \ $\chi'_F=\chi'$.

\smallskip\noindent
$\bullet$~Cas où $F\equiv\ov{X}(t_1,\ldots,t_k)$. On a alors~:\\
$\dsp\|p\force^fF\|=\{\tau\ps\pi;$ $\tau\in\C[p]\;,\pi\in\|\ov{X}(t_1,\ldots,t_k)\|\}$ et\\
$\dsp\|p\force'F\|=\bigcup_{q\in P}\{\pi^\tau;$
$\dsp\tau\in\C[pq],(\pi,q)\in\vv\ov{X}(t_1,\ldots,t_k)\vv\}=
\bigcup_{q\in P}$ $\{\pi^\tau;\;\;\tau\in\C[pq],\pi\in\|\ov{X}(t_1,\ldots,t_k)\|\}$.\\
On a donc \ $\chi_F=\lbd x(\chi)\lbd y(x)(\alpha_0)y$ \ et \ $\chi'_F=\lbd x\lbd y(\chi'x)(\alpha_4)y$ où $\alpha_0$
et $\alpha_4$ sont tels que~:\\
$\tau\in\C[pq]$ $\Fl$ $\alpha_0\tau\in\C[p]$~;
$\tau\in\C[p]$ $\Fl$ $\alpha_4\tau\in\C[p\1]$.

\cqfd

\smallskip\noindent
En appliquant le lemme~\ref{fforce'}, on en déduit~:
\begin{theorem}\label{forcef-fforce}\ \\
Pour chaque formule close $F[{\cal X}_1,\ldots,{\cal X}_k,\ov{Y}_1,\ldots,\ov{Y}_l]$ de SR$_1$,
à paramètres ${\cal X}_1,\ldots,{\cal X}_k$ dans SR$_1$ et $\ov{Y}_1,\ldots,\ov{Y}_l$ dans SR$_0$, il existe
deux quasi-preuves $\chi_F,\chi'_F$, qui ne dépendent pas des para\-mètres et des termes présents dans $F$,
telles que~:\\
$\xi\force(p\force^fF)$ \ $\Fl$ \ $(\chi_F\xi,p)\fforce F$ \ et \
$(\xi,p)\fforce F$ \ $\Fl$ \ $\chi'_F\xi\force(p\force^fF)$.
\end{theorem}\noindent

\begin{proposition}\label{p-pq-qp-forcef}\ \\
Soit $F[{\cal X}_1,\ldots,{\cal X}_k,\ov{Y}_1,\ldots,\ov{Y}_l]$ une formule close de SR$_1$, à paramètres
${\cal X}_1,\ldots,{\cal X}_k$ dans SR$_1$ et $\ov{Y}_1,\ldots,\ov{Y}_l$ dans SR$_0$. On a alors~:\\
$\lbd x(\chi'_F)(\ov{\gamma}_1)(\chi_F)x\force[(p\force^fF)\to(pq\force^fF)]$~;\\
$\lbd x(\chi'_F)(\ov{\gamma}_2)(\chi_F)x\force[(p\force^fF)\to(qp\force^fF)]$.
\end{proposition}\noindent
Si $\xi\force(p\force^fF)$, on a $((\chi_F)\xi,p)\fforce F$ (théorème~\ref{forcef-fforce}), donc
$((\ov{\gamma}_1)(\chi_F)\xi,pq)\fforce F$ (lemme\ref{force_le}).\\
D'où $(\chi'_F)(\ov{\gamma}_1)(\chi_F)\xi\force(pq\force^fF)$. On en déduit le premier résultat voulu.\\
Même preuve pour le second.

\cqfd

\begin{theorem}\label{forcef-hat}\ \\
Soit $F[\ov{X}_1,\ldots,\ov{X}_k]$ une formule close usuelle de SR$_0$. Il existe deux quasi-preuves closes $\chi^0_F,\chi^1_F$ telles que \ \
$(\chi^0_F,\chi^1_F)\force(p\force^fF[\ov{X}_1,\ldots,\ov{X}_k])\dbfl(\C[p]\to F[\ov{X}_1,\ldots,\ov{X}_k])$.
\end{theorem}\noindent
On a utilisé la notation \ $(\xi,\eta)\force A\dbfl B$ pour exprimer que l'on a
$\xi\force A\to B$ et $\eta\force B\to A$.

\smallskip\noindent
Preuve par récurrence sur $F$.\\
$\bullet$~~Si $F\equiv\ov{X}(t_1,\ldots,t_n)$, \ $p\force^fF$ est \ $\C[p]\to F$.

\smallskip\noindent
$\bullet$~~Si $F\equiv\pt x\,G$,  $p\force^fF$ est $\pt x(p\force^fG[n/x])$.
On peut donc prendre $\chi^0_F=\chi^0_G$ et $\chi^1_F=\chi^1_G$.

\smallskip\noindent
$\bullet$~~Si $F\equiv\pt X\,G$, même démonstration.

\smallskip\noindent
$\bullet$~~Si $F\equiv\pt x\indi G$, on a~:\\
$\dsp\|p\force^fF[\ov{X}_1,\ldots,\ov{X}_k]\|=
\bigcup_{n\in\NN}\|\{s^n0\}\to(p\force^fG[n,\ov{X}_1,\ldots,\ov{X}_k])\|$ et\\
$\dsp\|\C[p]\to F[\ov{X}_1,\ldots,\ov{X}_k]\|=
\bigcup_{n\in\NN}\|\C[p],\{s^n0\}\to G[n,\ov{X}_1,\ldots,\ov{X}_k]\|$.\\
On pose $\chi^0_F=\lbd x\lbd y\lbd z((\chi^0_G)(x)z)y$ et $\chi^1_F=\lbd x\lbd y(\chi^1_G)\lbd z(x)zy$.

\smallskip\noindent
$\bullet$~~Si $F\equiv G\to H$, on a \ 
$\dsp\|p\force^f(G\to H)\|=\bigcup_{q\in P}\|(q\force^fG)\to(pq\force^f H)\|$.\\
Donc, si $\alpha,\beta,\gamma$ sont des combinaisons des $\alpha_i(0\le i\le3)$ telles que~:\\
$\tau\in\C[pq]$ $\Fl$ $\alpha\tau\in\C[p]$, $\beta\tau\in\C[q]$ \ et
$\tau\in\C[p]$ $\Fl$ $\gamma\tau\in\C[pp]$, on peut poser\\
$\chi^0_F=\lbd x\lbd y\lbd z((\chi^0_H)(x)(\chi^1_G)\lbd d\,z)(\gamma)y$ \ et \
$\chi^1_F=\lbd x\lbd y(\chi^1_H)\lbd z((x)(\alpha)z)(\chi^0_Gy)(\beta)z$.\\
On le vérifie facilement en montrant que~:\\
$\chi^0_F\force[(p\force^fG[\ov{X}_1,\ldots,\ov{X}_k])\to(pp\force^f H[\ov{X}_1,\ldots,\ov{X}_k])],
\C[p],G[\ov{X}_1,\ldots,\ov{X}_k]\to H[\ov{X}_1,\ldots,\ov{X}_k]$\\
$\chi^1_F\force\{\C[p],G[\ov{X}_1,\ldots,\ov{X}_k]\to
H[\ov{X}_1,\ldots,\ov{X}_k]\},(q\force^fG[\ov{X}_1,\ldots,\ov{X}_k])
\to(pq\force^f H[\ov{X}_1,\ldots,\ov{X}_k])$.

\cqfd

\begin{theorem}\label{fforce-hat}\ \\
Soit $F[\ov{X}_1,\ldots,\ov{X}_k]$ une formule close usuelle de SR$_0$.
Il existe deux quasi-preuves closes $\chi^+_F,\chi^-_F$, qui ne
dépendent pas des paramètres de $F$, telles que~:\\
$(\xi,p)\fforce F[\ov{X}_1,\ldots,\ov{X}_n]$ \ $\Fl$ \
$\chi^+_F\xi\force\C[p]\to  F[\ov{X}_1,\ldots,\ov{X}_n]$~;\\
$\xi\force\C[p]\to  F[\ov{X}_1,\ldots,\ov{X}_n]$ \ $\Fl$ \
$(\chi^-_F\xi,p)\fforce F[\ov{X}_1,\ldots,\ov{X}_n]$.
\end{theorem}\noindent
Corollaire des théorèmes~\ref{forcef-fforce} et~\ref{forcef-hat}.

\cqfd

\subsection*{Le générique}
Dans la suite, on va considérer certaines fonctions $\phi:P^k\to P$. On considère alors une telle fonction
comme un symbole fonctionnel, qui permet donc de construire de nouveaux termes de condition (on a déjà
un symbole de fonction binaire et un symbole de constante $\1$). Bien entendu, l'interprétation du symbole
de fonction $\phi$ est cette fonction même.

\smallskip\noindent 
On ajoute au langage des formules de SR$_1$, un symbole $J$ \ de prédicat unaire \emph{sur les conditions} et le
quantificateur $\pt p^P$ sur les conditions. On définit donc les \emph{formules étendues de SR$_1$} comme suit~:

\smallskip\noindent
$\bullet$~~Les formules étendues {\em atomiques} de SR$_1$ sont

d'une part les formules atomiques de SR$_1$ ( de la forme  \ $t\ne u$, $X(t_1,\ldots,t_n)$,
$X^+(t_1,\ldots,t_n)$, où $X$ est une variable de prédicat d'arité $n$ et $t,u,t_1,\ldots,t_n$ sont des termes d'individu)~;

d'autre part les formules $J(t^{\PP})$, où $J$ est un symbole fixé (représentant, dans~SR$_1$,
\emph{l'idéal générique}) et $t^{\PP}$ est un terme de condition.\\
$\bullet$~~Si $F$ et $G$ sont des formules étendues de SR$_1$, alors $F\to G$ en est une aussi.\\
$\bullet$~~Si $G$ est une formule étendue de SR$_1$, alors $\C[t^{\PP}]\to G$ en est une aussi
($t^{\PP}$ un terme de condition).\\
$\bullet$~~Si $F$ est une formule étendue de SR$_1$, alors $\pt x\,F$, $\pt x\indi F$ et $\pt p^P\,F$
sont des formules étendues de~SR$_1$.\\
$\bullet$~~Si $F$ est une formule étendue de SR$_1$ et $X$ une variable de prédicat, alors $\pt X\,F$ et
$\pt X^+\,F$ sont des formules étendues de SR$_1$.

\smallskip\noindent
La valeur de vérité dans SR$_1$ d'une formule close étendue, avec paramètres dans SR$_1$ se définit comme
précédemment, par récurrence sur $F$. Il y a trois nouveaux cas pour la récurrence~:

\smallskip\noindent
$\bullet$~~Si $F$ est atomique close de la forme $J(t^{\PP})$, alors $t^{\PP}$ a une valeur $p\in P$. On pose~:\\
\centerline{$\vv J(t^{\PP})\vv=\Pi\fois\{p\}$.}

\noindent
$\bullet$~~Si $F\;\equiv\;\C[t^{\PP}]\to G$, alors $t^{\PP}$ a une valeur $p\in P$. On pose~:\\
\centerline{$\vv F\vv=\{(\tau,\1)\ps(\pi,q);$ $\tau\in\C[p],(\pi,q)\in\vv G\vv\}$.}

\noindent
$\bullet$~~Si $F\;\equiv\;\pt p^PG$, alors \ $\dsp\vv F\vv=\bigcup_{p\in P}\vv G[p]\vv$.

\smallskip\noindent
Corrélativement, on ajoute au langage de SR$_0$, les formules atomiques \ $t^{\PP}\neps J(u^{\PP})$,
où $t^{\PP},u^{\PP}$ sont des termes de condition.\\
Si une telle formule atomique est close, alors $t^{\PP},u^{\PP}$ prennent respectivement les valeurs
$p,q\in P$. Donc $\|t^{\PP}\neps J(u^{\PP})\|=\{\pi\in\Pi;$ $(\pi,p)\in\Pi\fois\{q\}\}=\Pi$ si $p=q$
et $=\vide$ si $p\ne q$.\\
Autrement dit, la formule atomique \ $t^{\PP}\neps J(u^{\PP})$ pourra être écrite plus simplement
${t^{\PP}\ne u^{\PP}}$.

\smallskip\noindent
On peut alors définir la formule \ $p\force^fF$ de SR$_0$, pour toute formule $F$ étendue de SR$_1$.
Dans la définition par récurrence sur $F$, il y a deux nouveaux cas~:\\
$\bullet$~~Si $F$ est atomique de la forme $J(t^{\PP})$, on pose~:\\
$p\force^fF\;\equiv\;\pt r^P(\C[pr]\to r\neps J(t[p_1,\ldots,p_k]))$.\\
Comme la formule $r\neps J(t^{\PP})$ n'est autre que \ $r\ne t^{\PP}$, on voit que~:\\
$p\force^f J(t^{\PP})\;\equiv\;\neg\C[pt^{\PP}]$.\\
$\bullet$~~Si $F\equiv\pt p^PG$, alors $p\force^fF\equiv\pt p^P(p\force^fG)$.

\begin{proposition}\label{fforce_Jp}\ \\
i) Dans les formules de SR$_0$, on a~:\\
$p\neps J(q)\;\equiv\;p\neq q$~; \ $p\force^fJ(q)\;\equiv\;\neg\C[pq]$.\\
ii) $(\xi,p)\fforce J(q)$ $\Fl$ $\chi'\xi\force\neg\C[pq]$~; \
$\xi\force\neg\C[pq]$ $\Fl$ $(\chi\xi,p)\fforce J(q)$.
\end{proposition}\noindent
i) a déjà été montré.\\
ii) Si $(\xi,p)\fforce J(q)$, on a $(\xi,p)\star(\pi,q)\in\bbot$ pour tout $\pi\in\Pi$. Donc~:\\
$\tau\in\C[pq]$ $\Fl$ $\xi\star\pi^\tau\in\bbot$ $\Fl$ $\chi'\xi\star\tau\ps\pi\in\bbot$.\\
Inversement, si $\xi\force\neg\C[pq]$, alors $\xi\star\tau\ps\pi\in\bbot$ pour tout $\tau\in\C[pq]$, donc
$\chi\xi\star\pi^\tau\in\bbot$, d'où $(\chi\xi,p)\star(\pi,q)\in\bbbot$.

\cqfd

\smallskip\noindent
Le théorème~\ref{forcef-fforce} reste donc vrai pour le langage étendu. Il suffit, en effet, de le vérifier pour
les nouvelles formules atomiques $J(p)$ du langage de SR$_1$, ce qui résulte immédiatement de la
proposition~\ref{fforce_Jp}.

\smallskip\noindent
{\bfseries Notations.}\\
$\bullet$~~On note $p\le q$ la formule $\pt r^P(\neg\C[qr]\to\neg\C[pr])$ \ et \
$p\sim q$ la formule $p\le q\land q\le p$, c'est-à-dire \ $\pt r^P(\neg\C[qr]\dbfl\neg\C[pr])$.\\
Dans la suite, on écrira souvent $F\to\C[p]$ au lieu de $\neg\C[p]\to\neg F$~;\\
$p\le q$ s'écrit alors $\pt r^P(\C[pr]\to\C[qr])$ et $p\sim q$ s'écrit $\pt r^P(\C[pr]\dbfl\C[qr])$.\\
{\small{\bfseries Remarque.} On rappelle, en effet, que $\C[p]$ n'est pas une formule, mais une partie de $\Lbd_c$~;
en fait, dans les structures de réalisabilité considérées plus loin, il existera une formule $\CC[p]$ de SR$_0$
telle que $\C[p]=\{\tau\in\Lbd_c;$ $\tau\force\CC[p]\}$.
On pourra alors identifier $\C[p]$ à la formule $\CC[p]$.}\\
$\bullet$~~Si $F$ est une formule close de SR$_1$, on écrira \ $\fforce F$ \ pour exprimer qu'il existe
une quasi-preuve $\theta$ telle que \ $(\theta,\1)\fforce F$. D'après la proposition~\ref{1_to_p},
cela équivaut à dire qu'il existe une quasi-preuve $\theta$ telle que \ $(\theta,p)\fforce F$ pour
tout $p\in P$.

\begin{proposition}\label{1_to_p}
Si \ $(\theta,\1)\fforce F$, alors \ $(\ov{\alpha}\theta,p)\fforce F$ pour tout $p\in P$~; \ $\alpha$ est une
quasi-preuve telle que \ $\tau\in\C[pq]$ $\Fl$ $\alpha\tau\in\C[\1q]$.
\end{proposition}\noindent
En effet, on doit montrer que, pour tout $(\pi,q)\in\vv F\vv$, on a \
$(\ov{\alpha}\theta,p)\star(\pi,q)\in\bbbot$, soit~:\\
$(\ov{\alpha}\theta\star\pi,pq)\in\bbbot$. Soit donc $\tau\in\C[pq]$, d'où $\alpha\tau\in\C[\1q]$.\\
Comme on a, par hypothèse, $(\theta\star\pi,\1q)\in\bbbot$, on en déduit \
$\theta\star\pi^{\alpha\tau}\in\bbot$ et donc $\ov{\alpha}\theta\star\pi^\tau\in\bbot$.

\cqfd

\begin{theorem}[Propriétés élémentaires du générique]\label{elem_gen}\ \\
i) $(\ov{\alpha},r)\fforce\neg J(\1)$ pour tout $r\in P$\\
où $\alpha$ une quasi-preuve telle que $\tau\in\C[r(pq)]$ $\Fl$ $\alpha\tau\in\C[p\1]$.\\
ii) $(\theta,r)\fforce\pt p^P(\neg\C[p]\to J(p))$, quel que soit $r\in P$\\
où $\theta=\chi\lbd x\lbd y((\chi'y)(\beta)x)(\alpha)x$\\
et \ $\alpha,\beta$ sont deux quasi-preuves telles que $\tau\in\C[r(qp)]$ $\Fl$ $\alpha\tau\in\C[p]$
et $\beta\tau\in\C[q(\1r)]$.\\
iii) $(\theta,r)\fforce\neg J(p),J(pq)\to J(q)$ pour tous $p,q,r\in P$\\
où $\theta=\lbd x\lbd y(\ov{\alpha})(x)(\ov{\beta})y$ \ et \
$\alpha,\beta$ sont deux quasi-preuves telles que~:\\
$\tau\in\C[r(p'(q'q))]$ $\Fl$ $\alpha\tau\in\C[p'((qq')\1)]$ \ et \
$\tau\in\C[(qq')p]$ $\Fl$ $\beta\tau\in\C[q'(pq)]$.\\
iv) $\fforce\pt p^P(\pt q^P(\neg\C[pq]\to J(q))\to\neg J(p))$.\\
v) $\fforce\pt p^P\pt q^P(J(p),q\le p\to J(q))$.\\
vi) $\fforce\pt p^P\pt q^P(\neg\C[pq],\neg J(p)\to J(q))$.
\end{theorem}\noindent
i) Soit $(\xi,p)\fforce J(\1)$~; on doit montrer que \ $(\ov{\alpha},r)\star(\xi,p)\ps(\pi,q)\in\bbbot$, soit ${(\ov{\alpha}\star\xi\ps\pi,r(pq))\in\bbbot}$.
Soit donc $\tau\in\C[r(pq)]$~; on doit montrer que \ $\ov{\alpha}\star\xi\ps\pi^\tau\in\bbot$, soit $\xi\star\pi^{\alpha\tau}\in\bbot$. Mais on a
$\alpha\tau\in\C[p\1]$ et, par hypothèse sur $(\xi,p)$, on a $(\xi,p)\star(\pi,\1)\in\bbbot$, donc
$(\xi\star\pi,p\1)\in\bbbot$, d'où le résultat.

\smallskip\noindent
ii)Soient $(\eta,q)\fforce\neg\C[p]$ et $(\pi,p)\in\vv J(p)\vv$. On doit montrer que
$(\theta,r)\star(\eta,q)\ps(\pi,p)\in\bbbot$, soit $(\theta\star\eta\ps\pi,r(qp))\in\bbbot$.
Soit donc $\tau\in\C[r(qp)]$~; on doit montrer que $\theta\star\eta\ps\pi^\tau\in\bbot$.\\
Or, on a $\alpha\tau\in\C[p]$~; par hypothèse sur $(\eta,q)$, on a donc
$(\eta,q)\star(\alpha\tau,\1)\ps(\pi,r)\in\bbot$, soit~:\\
$(\eta\star\alpha\tau\ps\pi,q(\1r))\in\bbot$. Mais on a $\beta\tau\in\C[q(\1r)]$, donc $\eta\star\alpha\tau\ps\pi^{\beta\tau}\in\bbot$, d'où $\theta\star\eta\ps\pi^\tau\in\bbot$.

\smallskip\noindent
iii) Soient $(\xi,p')\fforce\neg J(p)$, \ $(\eta,q')\fforce J(pq)$ \ et \ $(\pi,q)\in\vv J(q)\vv$.
On doit montrer que~:\\
$(\theta,r)\star(\xi,p')\ps(\eta,q')\ps(\pi,q)\in\bbot$, soit \
$(\theta\star\xi\ps\eta\ps\pi,r(p'(q'q)))\in\bbot$.\\
Soit donc $\tau\in\C[r(p'(q'q))]$~; on doit montrer que \ $\theta\star\xi\ps\eta\ps\pi^\tau\in\bbot$.\\
On montre d'abord que \ $(\ov{\beta}\eta,qq')\fforce J(p)$, c'est-à-dire \
$(\ov{\beta}\eta,qq')\star(\varpi,p)\in\bbbot$ pour toute $\varpi\in\Pi$.\\
Ceci s'écrit \ $(\ov{\beta}\eta\star\varpi,(qq')p)\in\bbbot$. Soit donc $\upsilon\in\C[(qq')p]$~;
on a $\beta\upsilon\in\C[q'(pq)]$~; or, par hypothèse sur $\eta$, on a $(\eta,q')\star(\varpi,pq)\in\bbbot$,
donc $(\eta\star\varpi,q'(pq))\in\bbbot$, d'où $\eta\star\varpi^{\beta\upsilon}\in\bbot$ et, finalement
$\ov{\beta}\eta\star\varpi^\tau\in\bbot$, ce qui est le résultat voulu.\\
De  \ $(\ov{\beta}\eta,qq')\fforce J(p)$, on déduit, par hypothèse sur $\xi$, que \
$(\xi,p')\star(\ov{\beta}\eta,qq')\ps(\pi,\1)\in\bbbot$, soit~:\\
$(\xi\star\ov{\beta}\eta\ps\pi,p'((qq')\1))\in\bbbot$. Or, on a \ $\alpha\tau\in\C[p'((qq')\1)]$, donc \
$\xi\star\ov{\beta}\eta\ps\pi^{\alpha\tau}\in\bbot$, \ d'où~:\\
$\ov{\alpha}\star(\xi)(\ov{\beta})\eta\ps\pi^\tau\in\bbot$ \ et, enfin \
$\theta\star\xi\ps\eta\ps\pi^\tau\in\bbot$.

\smallskip\noindent
iv) Soient $(\xi,q)\fforce J(p)$ et $(\eta,r)\fforce\pt q(\neg\C[pq]\to J(q))$~; on cherche une quasi-preuve $\theta$
telle que \ $(\theta,\1)\star(\eta,r)\ps(\xi,q)\ps(\pi,r')\in\bbbot$, soit
$(\theta\star\eta\ps\xi\ps\pi,\1(r(qr')))\in\bbbot$.\\
D'après la proposition~\ref{fforce_Jp}, on a $\chi'\xi\force\neg\C[pq]$, donc
$\lbd d\chi'\xi\force\C[\1]\to\neg\C[pq]$.\\
D'après le théorème~\ref{fforce-hat}, il existe donc une
quasi-preuve $\chi_0$ telle que $(\chi_0\xi,\1)\fforce\neg\C[pq]$.\\
Par hypothèse sur $\eta$, on a donc $(\eta,r)\star(\chi_0\xi,\1)\ps(\pi,q)\in\bbbot$, soit
$(\eta\star\chi_0\xi\ps\pi,r(\1q))\in\bbbot$. On peut donc prendre
$\theta=(\ov{\alpha})\lbd x\lbd y(y)(\chi_0)x$ \ où $\alpha$ est une quasi-preuve telle que~:\\
$\tau\in\C[\1(r(qr'))]$ $\Fl$ $\alpha\tau\in\C[r(\1q)]$.\\
v) Soient $(\xi,p')\fforce J(p)$ et $(\eta,r)\fforce q\le p$~; on cherche une quasi-preuve $\theta$ telle que~:\\
$(\theta,\1)\star(\xi,p')\ps(\eta,r)\ps(\pi,q)\in\bbbot$ pour toute $\pi\in\Pi$, soit
$(\theta\star\xi\ps\eta\ps\pi,\1(p'(rq)))\in\bbbot$.\\
D'après la proposition~\ref{fforce_Jp}, on a $\chi'\xi\force\neg\C[pp']$.
D'après le théorème~\ref{fforce-hat}, il existe une
quasi-preuve $\chi_0$ telle que $\chi_0\eta\force\C[r]\to q\le p$. Or, on a facilement~:\\
$\vdash\neg\C[pp'],(\C[r]\to q\le p)\to\neg\C[(p'r)q]$.\\
Il existe donc une quasi-preuve $\chi_1$ telle que $\chi_1\xi\eta\force\neg\C[(p'r)q]$~;
d'après la proposition~\ref{fforce_Jp}, on a donc $(\chi_2\xi\eta,p'r)\fforce J(q)$, pour une quasi-preuve
$\chi_2$ convenable. On a donc~:\\
$(\chi_2\xi\eta,p'r)\star(\pi,q)\in\bbbot$, soit \ $(\chi_2\xi\eta\star\pi,(p'r)q)\in\bbbot$. On peut donc
prendre $\theta=\ov{\alpha}\chi_2$, où $\alpha$ est une quasi-preuve telle que \
$\tau\in\C[\1(p'(rq))]$ $\Fl$ $\alpha\tau\in\C[(p'r)q]$.\\
vi) Soient $(\xi,r)\fforce\neg\C[pq]$ et $(\eta,q')\fforce\neg J(p)$~; on cherche une quasi-preuve
$\theta$ telle que~:\\
$(\theta,\1)\star(\xi,r)\ps(\eta,q')\ps(\pi,q)\in\bbbot$ pour toute $\pi\in\Pi$, soit
$(\theta\star\xi\ps\eta\ps\pi,\1(r(q'q)))\in\bbbot$.\\
D'après le théorème~\ref{fforce-hat}, il existe une quasi-preuve $\chi_0$ telle que $\chi_0\xi\force\C[r]\to\neg\C[pq]$.\\
Or, on a facilement \ $\vdash(\C[r]\to\neg\C[pq])\to\neg\C[(qr)p]$.\\
Il existe donc une quasi-preuve $\chi_1$ telle que $\chi_1\xi\force\neg\C[(qr)p]$~;
d'après la proposition~\ref{fforce_Jp}, on a donc $(\chi_2\xi,qr)\fforce J(p)$, pour une quasi-preuve
$\chi_2$ convenable. On a donc~:\\
$(\eta,q')\star(\chi_2\xi,qr)\ps(\pi,\1)\in\bbbot$, par hypothèse sur $\eta$~; ce qui s'écrit \
${(\eta\star\chi_2\xi\ps\pi,q'((qr)\1))\in\bbbot}$.\\
On cherche une quasi-preuve $\theta$ telle que $\theta\star\xi\ps\eta\ps\pi^\tau\in\bbot$, pour tout
$\tau\in\C[\1(r(q'q))]$.
Soit $\alpha$ une quasi-preuve telle que \ $\tau\in\C[\1(r(q'q))]$ $\Fl$
$\alpha\tau\in\C[q'((qr)\1)]$. On a donc $\eta\star\chi_2\xi\ps\pi^{\alpha\tau}\in\bbot$,
donc $\ov{\alpha}\star\eta\ps\chi_2\xi\ps\pi^\tau\in\bbot$, ce qui montre qu'on peut prendre
$\theta=\lbd x\lbd y(\ov{\alpha}y)(\chi_2)x$.

\cqfd

\begin{theorem}[Densité]\label{densite}\ \\
Pour toute fonction $\phi~:P\to P$, on a~:\\
$(\theta,\1)\fforce\pt p^P(\neg\C[p\phi(p)]\to J(p)),\pt p^P\,J(p\phi(p))\to\bot$\\
avec $\theta=(\ov{\beta})\lbd x\lbd y(x)(\vartheta)y$, \
$\vartheta=(\chi)\lbd d\lbd x\lbd y(\chi'x)(\alpha)y$~; \
$\alpha,\beta$ sont deux quasi-preuves telles~que~:
$\tau\in\C[qr]$ $\Fl$ $\alpha\tau\in\C[q(qr)]$~; \
$\tau\in\C[\1(p(qr))])$ $\Fl$ $\beta\tau\in\C[p(\1q)]$.
\end{theorem}\noindent
Soient $(\xi,p_0)\fforce\pt p^P(\neg\C[p\phi(p)]\to J(p))$, $(\eta,q_0)\fforce\pt p^P\,J(p\phi(p))$
et $(\pi,r_0)\in\PPi$.\\
On doit montrer que $(\theta\star\xi\ps\eta\ps\pi,\1(p_0(q_0r_0)))\in\bbbot$, c'est-à-dire~:\\
$(*)$\hspace{6em}$(\pt\tau\in\C[\1(p_0(q_0r_0))])\,\theta\star\xi\ps\eta\ps\pi^\tau\in\bbot$.

\smallskip\noindent
On montre d'abord que $(\vartheta\eta,\1)\fforce\neg\C[q_0\phi(q_0)]$.\\
Pour cela, on doit montrer \ $(\pt(\varpi,r)\in\PPi)(\pt\tau\in\C[q_0\phi(q_0)])
(\vartheta\eta,\1)\star(\tau,\1)\ps(\varpi,r)\in\bbbot$\\
ou encore \ $(\pt\varpi\in\Pi)(\pt r\in P)(\pt\tau\in\C[q_0\phi(q_0)])(\pt\tau'\in\C[\1(\1r)])
\,\vartheta\eta\star\tau\ps\varpi^{\tau'}\in\bbot$, c'est-à-dire~:\\
$(**)$\hspace{5em}$(\pt\varpi\in\Pi)(\pt\tau\in\C[q_0\phi(q_0)])\,\eta\star\varpi^{\alpha\tau}\in\bbot$.\\
Or, par hypothèse sur $\eta$, on a $(\eta,q_0)\fforce J(q_0\phi(q_0))$, c'est-à-dire~:\\
$(\eta\star\varpi,q_0(q_0\phi(q_0)))\in\bbbot$ ou encore \
$(\pt\tau\in\C[q_0(q_0\phi(q_0))])\,\eta\star\varpi^\tau\in\bbot$.\\
Cela donne le résultat $(**)$ cherché.

\smallskip\noindent
Par hypothèse sur $\xi$, on a $(\xi,p_0)\fforce\neg\C[q_0\phi(q_0)]\to J(q_0)$. Il en résulte que~:\\
$(\pt\pi\in\Pi)(\xi,p_0)\star(\vartheta\eta,\1)\ps(\pi,q_0)\in\bbbot$, soit \
$(\pt\pi\in\Pi)(\xi\star\vartheta\eta\ps\pi,p_0(\1q_0))\in\bbbot$.\\
On en déduit \ $(\pt\pi\in\Pi)(\pt\tau\in\C[p_0(\1q_0)])\xi\star\vartheta\eta\ps\pi^\tau\in\bbot$.\\
Cela donne le résultat voulu, puisque l'on a \
$\theta\star\xi\ps\eta\ps\pi^\tau\succ\xi\star\vartheta\eta\ps\pi^{\beta\tau}$\\
et $\tau\in\C[\1(p_0(q_0r_0))])$ $\Fl$ $\beta\tau\in\C[p_0(\1q_0)]$.

\cqfd

\subsection*{Condition de chaîne dénombrable}
{\bfseries Notations.} La formule $\pt x\indi(Xx\to Yx)$ où $X,Y$ sont des variables de prédicat unaire, sera notée $X\subseteq Y$. La formule $(X\subseteq Y)\land(Y\subseteq X)$ sera notée $X\simeq Y$.\\
Même chose, en remplaçant $X,Y$ par $X^+,Y^+$ (avec les quatre combinaisons possibles).

\noindent
Rappelons que, si ${\cal X}:\NN\fois P\to{\cal P}(\Pi)$ est un prédicat unaire de SR$_1$, alors ${\cal X}(n,p)$
est aussi écrit $\|p\neps{\cal X}n\|$. La formule \ $\pt q^P(\C[pq]\to q\neps{\cal X}n)$ est écrite
$p\force^f{\cal X}n$.

\smallskip\noindent
On dira que \ $\C$ \ satisfait la \emph{condition de chaîne dénombrable} s'il existe une application\\
$\p:{\cal P}(\Pi)^{\NN\fois P}\to P$ et deux quasi-preuves ${\sf cd}_0,{\sf cd}_1$ telles que, pour tout
${\cal X}\in{\cal P}(\Pi)^{\NN\fois P}$~:\\
${\sf cd}_0\force\pt n\indi\pt q^P(q\neps{\cal X}n\to\p({\cal X})\le q)$~;\\
${\sf cd}_1\force
\pt n\indi\pt q^P\pt r^P(q\neps{\cal X}n,r\neps{\cal X}n\to q=r),\pt n\indi\ex q^P(q\neps{\cal X}n),\\
\pt n\indi\pt q^P(q\neps{\cal X}n\to\C[q]),
\pt m\indi\pt n\indi\pt q^P\pt r^P(q\neps{\cal X}n,r\neps{\cal X}(n+m)\to r\le q)
\to\C[\p({\cal X})]$.

\smallskip\noindent
{\small{\bfseries Remarques.}\\
$\bullet$~~Rappels~: $\C[p]$ n'est pas une formule, mais un ensemble de $\lbd_c$-termes~;
la notation $F\to\C[p]$ désigne la formule $F,\neg\C[p]\to\bot$~; \
$p\le q$ est la formule $\pt r^P(\C[pr]\to\C[qr])$.\\
$\bullet$~~Intuitivement, on considère ${\cal X}n$ comme une partie de $P$~: c'est l'ensemble
des conditions $p$ telles que $p\neps{\cal X}n$. Cette formule exprime donc que~:\\
$\p({\cal X})$ est une condition plus forte que toutes celles des ${\cal X}n$\\
(traduction de $\pt n\indi\pt q^P(q\neps{\cal X}n\to\p({\cal X})\le q)$)~;\\ 
si chaque ${\cal X}n$ \ a exactement un élément $p_n$\\
(traduction de $\pt n\indi\pt q^P\pt r^P(q\neps{\cal X}n,r\neps{\cal X}n\to q=r)$ et
$\pt n\indi\ex q^P(q\neps{\cal X}n)$),\\
si la suite $p_n$ est décroissante\\
(traduction de $\pt m\indi\pt n\indi\pt q^P\pt r^P(q\neps{\cal X}n,r\neps{\cal X}(n+m)\to r\le q)$),\\
si $p_n$ est une condition non triviale pour tout $n$\\
(traduction de $\pt n\indi\pt q^P(q\neps{\cal X}n\to\C[q])$),\\
alors $\p({\cal X})$ est une condition non triviale\\
(traduction de $\C[\p({\cal X})]$).}

\smallskip\noindent
{\bfseries Notation.} On désigne par \ $p\force^f\pm{\cal X}n$ \ la formule
$\pt q^P(\C[pq],q\force^f{\cal X}n\to p\force^f{\cal X}n)$ \ ou encore \
$\pt q^P\pt r^P(\C[pq],\C[pr],q\force^f{\cal X}n\to r\neps{\cal X}n)$, \
qui se lit \gmg la condition $p$ \ décide ${\cal X}n$\gmd.

\begin{theorem}\label{inf_chaine_den}
Si $\C$ a la propriété de chaîne dénombrable, alors il existe une application
$(p,{\cal X})\mapsto p^{\cal X}$ de $P\fois{\cal P}(\Pi)^{\NN\fois P}$ dans $P$ et des quasi-preuves
$\delta_0,\delta_1,\delta_2$ telles que~:\\
$\delta_0\force\C[p]\to\C[p^{\cal X}]$~; \ $\delta_1\force p^{\cal X}\le p$~; \
$\delta_2\force\pt n\indi(p^{\cal X}\force^f\pm{\cal X}n)$.
\end{theorem}\noindent
{\small{\bfseries Remarque.} Ces propriétés se traduisent par~: \ \gmg si $p$ est une condition non triviale,
alors $p^{\cal X}$ est une condition non triviale, plus forte que $p$, qui décide ${\cal X}n$ pour tout
entier $n$\gmd.}

\begin{lemma}\label{signature}
Il existe une fonction $\phi:\NN\fois P\fois{\cal P}(\Pi\fois P)\to P$ telle que~:\\
(1)\hspace{6em}$\sigma\force\pt x\indi F(p,\phi(x,p,{\cal X}n),{\cal X}n)\to p\force^f{\cal X}n$\\
où \ $F(p,q,{\cal X}n)\;\equiv\;\C[pq]\to q\neps{\cal X}n$, donc \
$p\force^f{\cal X}n\;\equiv\;\pt q^P\,F(p,q,{\cal X}n)$.
\end{lemma}\noindent
C'est la preuve usuelle que $\sigma$ réalise l'axiome du choix non extensionnel (ACNE)[3,4,5].\\
La règle de réduction pour l'instruction $\sigma$ (signature) est la suivante~:\\
\centerline{$\sigma\star\xi\ps\pi\succ\xi\star s^j\ul{0}\ps\pi$}

\noindent
où $j$ est le numéro de la pile $\pi$ dans une énumération récursive $i\mapsto\pi_i$ fixée de $\Pi$.\\
On définit $\phi$ au moyen de l'axiome du choix, de façon que, pour tout $i\in\NN$, on ait~:\\
\centerline{$\dsp\pi_i\in\bigcup_{q\in P}\|F(p,q,{\cal X}n)\|$ \ $\Fl$ \
$\pi_i\in\|F(p,\phi(i,p,{\cal X}n),{\cal X}n)\|$.}

\smallskip\noindent
Soient alors $\xi\force\pt x\indi F(p,\phi(x,p,{\cal X}n),{\cal X}n)$ et $\pi\in\|p\force^f{\cal X}n\|$.\\
On a $\pi=\pi_j$ pour un entier $j$ et, par ailleurs,
$\dsp\|p\force^f{\cal X}n\|=\bigcup_{q\in P}\|F(p,q,{\cal X}n)\|$.\\
On a donc $\pi\in\|F(p,\phi(j,p,{\cal X}n),{\cal X}n)\|$.\\
On doit montrer que $\sigma\star\xi\ps\pi\in\bbot$, soit $\xi\star s^j\ul{0}\ps\pi\in\bbot$. Or, on a~:

\smallskip\noindent
$\dsp s^j\ul{0}\ps\pi\in\|\{s^j\ul{0}\}\to F(p,\phi(j,p,{\cal X}n),{\cal X}n)\|\subset
\bigcup_{i\in\NN}\|\{s^i\ul{0}\}\to F(p,\phi(i,p,{\cal X}n),{\cal X}n)\|\\
=\|\pt x\indi F(p,\phi(x,p,{\cal X}n),{\cal X}n)\|$.
D'où le résultat, par hypothèse sur $\xi$.

\cqfd

\smallskip\noindent
On définit ensuite la formule $\Psi(r,p,{\cal X}n)$ de SR$_0$~:\\
$\Psi(r,p,{\cal X}n)\;\equiv\;(p\force^f{\cal X}n\to r=p)\land\\
\pt x\indi[\neg F(p,\phi(x,p,{\cal X}n),{\cal X}n),
\pt y\indi\pt z\indi(y+z+1=x\to F(p,\phi(y,p,{\cal X}n),{\cal X}n))\\
\hspace*{\fill}\to r=p\phi(x,p,{\cal X}n)]$.

\smallskip\noindent
La formule $\Psi(r,p,{\cal X}n)$ exprime que l'on a \ $r=\psi(p,{\cal X}n)$, où~:\\
$\psi(p,{\cal X}n)=p$ si $p\force^f{\cal X}n$, et sinon\\
$\psi(p,{\cal X}n)=p\phi(x,p,{\cal X}n)$ pour le premier entier $x$ tel que l'on ait \ $\neg F(p,\phi(x,p,{\cal X}n),{\cal X}n)$.
Un tel entier existe toujours, d'après (1).\\
Noter que $\psi$, contrairement à $\phi$, \emph{n'est pas} un symbole de fonction, c'est-à-dire une application
de $P\fois{\cal P}(\Pi\fois P)$ dans $P$. On ne peut l'utiliser que par l'intermédiaire de la formule
$r=\psi(p,{\cal X}n)$, qui s'écrit $\Psi(r,p,{\cal X}n)$.

\begin{lemma}\label{psi_decide}\ \\
i) $\force\pt n\indi\ex r^P\Psi(r,p,{\cal X}n)$~;
$\force\pt n\indi\pt q^P\pt r^P(\Psi(q,p,{\cal X}n),\Psi(r,p,{\cal X}n)\to q=r)$~;\\
ii) $\force\pt r^P(\Psi(r,p,{\cal X}n)\to r\le p)$~;\\
iii) $\force\pt r^P(\Psi(r,p,{\cal X}n),\C[p]\to\C[r])$~;\\
iv) $\force\pt r^P(\Psi(r,p,{\cal X}n)\to(r\force^f\pm{\cal X}n))$.
\end{lemma}\noindent
{\small{\bfseries Remarque.} Rappelons que la notation $\force A(p,{\cal X}n)$ signifie qu'il existe une
quasi-preuve, indépendante de $p,{\cal X},n$, qui réalise $A(p,{\cal X}n)$.}

\smallskip\noindent
Il suffit de prouver ces formules à l'aide d'hypothèses déjà réalisées par des quasi-preuves.\\
Les formules (i) expriment que $\Psi(r,p,{\cal X}n)$ est fonctionnelle, ce qui est conséquence
de sa définition et du lemme~\ref{signature} (formule~1).\\
On prouve maintenant (ii), (iii) et (iv) sous la forme~:\\
$\psi(p,{\cal X}n)\le p$, \ $\C[p]\to\C[\psi(p,{\cal X}n)]$ \ et $\psi(p,{\cal X}n)\force^f\pm{\cal X}n$.
On distingue deux cas.

\smallskip\noindent
$\bullet$~~Si $p\force^f{\cal X}n$, alors on a $\psi(p,{\cal X}n)=p$, donc $\psi(p,{\cal X}n)\le p$ et
$\C[p]\to\C[\psi(p,{\cal X}n)]$~;\\
de plus, puisqu'on a $p\force^f{\cal X}n$, on a trivialement $p\force^f\pm{\cal X}n$ et donc
$\psi(p,{\cal X}n)\force^f\pm{\cal X}n$.\\
$\bullet$~~Si $p\nnforce^f{\cal X}n$, alors on a $\psi(p,{\cal X}n)=pq$, avec $q=\phi(x,p,{\cal X}n)$
pour un certain entier $x$~;\\
on en déduit $\psi(p,{\cal X}n)\le p$.\\
On a, de plus, $\neg F(p,q,{\cal X}n)$, autrement dit $\C[pq]\land q\eps{\cal X}n$.\\
$\C[pq]$ s'écrit $\C[\psi(p,{\cal X}n)]$, d'où l'on déduit trivialement $\C[p]\to\C[\psi(p,{\cal X}n)]$.
D'autre part, de $q\eps{\cal X}n$, on déduit $\pt r^P(\C[qr],r\force^f{\cal X}n\to\bot)$, d'où $q\force^f\pm{\cal X}n$
et donc $pq\force^f\pm{\cal X}n$, c'est-à-dire $\psi(p,{\cal X}n)\force^f\pm{\cal X}n$.

\cqfd

\smallskip\noindent
On définit alors, par récurrence sur $n$, le prédicat unaire ${\cal Y}:\NN\fois P\to{\cal P}(\Pi)$
de SR$_1$ en posant~:
${\cal Y}(n,q)(=\|q\neps{\cal Y}n\|)=\|q=p_n\|$ avec $p_0=p$ et $p_{n+1}=\psi(p_n,{\cal X}n)$.\\
En première approximation (c'est-à-dire en utilisant le symbole auxiliaire de fonction $\psi$),
la formule pour \ $q\neps{\cal Y}n$ est donc~:\\
(2)\hspace{2em}$q\neps{\cal Y}n\;\equiv\;\pt Z^+\{\pt j\indi\pt r^P(r\neps Z^+j\to\psi(r,{\cal X}j)\neps Z^+(j+1)),
p\neps Z^+0\to q\neps Z^+n\}$.

\smallskip\noindent
Son écriture exacte est donc la formule \ $\Phi(n,q,p,{\cal X})$~:\\
$\pt Z^+\{\pt j\indi\pt r^P\pt r'^P(r\neps Z^+j,\Psi(r',r,{\cal X}j)\to r'\neps Z^+(j+1)),
p\neps Z^+0\to q\neps Z^+n\}$.\\
On définit donc la fonction ${\cal Y}:\NN\fois P\to{\cal P}(\Pi)$ en posant 
$\|q\neps{\cal Y}n\|={\cal Y}(n,q)=\|\Phi(n,q,p,{\cal X})\|$.

\begin{lemma}\label{suite_cal_Y}\ \\
Les formules suivantes sont réalisées par des quasi-preuves indépendantes de $p,{\cal X}$~:\\
i)~~$\pt q^P(q\neps{\cal Y}0\dbfl q=p)$~;\\
ii)~~$\pt n\indi\ex q^P(q\neps{\cal Y}n)$~;\\
iii~~$\pt n\indi\pt q^P\pt r^P(q\neps{\cal Y}n,r\neps{\cal Y}n\to q=r)$~;\\
iv)~~$\pt m\indi\pt n\indi\pt q^P\pt r^P(q\neps{\cal Y}n,r\neps{\cal Y}(n+m)\to r\le q)$~;\\
v)~~$\pt n\indi\pt q^P(q\neps{\cal Y}(n+1)\to q\force^f\pm{\cal X}n)$~;\\
vi)~~$\C[p]\to\pt n\indi\pt q^P(q\neps{\cal Y}n\to\C[q])$.
\end{lemma}\noindent
Il suffit de montrer que ces formules sont conséquences de formules réalisées par des quasi-preuves et de la
définition de $q\neps{\cal Y}n$ (on utilisera l'expression (2)).

\smallskip\noindent
On montre d'abord~:\\
(3)\hspace{5em}$\force p\neps{\cal Y}0$ et
$\force\pt j\indi\pt r^P(r\neps{\cal Y}j\to\psi(r,{\cal X}j)\neps{\cal Y}(j+1))$.\\
Le premier résultat découle trivialement de la définition de $p\neps{\cal Y}0$.\\
$r\neps{\cal Y}i$ s'écrit $\pt Z^+\{\pt j\indi\pt r^P(r\neps Z^+j\to\psi(r,{\cal X}j)\neps Z^+(j+1)),
p\neps Z^+0\to r\neps Z^+i\}$.\\
On en déduit~:\\
$\pt Z^+\{\pt j\indi\pt r^P(r\neps Z^+j\to\psi(r,{\cal X}j)\neps Z^+(j+1)),p\neps Z^+0
\to \psi(r,{\cal X}i)\neps Z^+(i+1)\}$,\\
c'est-à-dire $\psi(r,{\cal X}i)\neps{\cal Y}(i+1)$. C'est le deuxième résultat de (3).

\smallskip\noindent
De~(3), on déduit immédiatement $\pt n\indi\ex q^P(q\neps{\cal Y}n)$ par récurrence sur $n$, ce qui donne (ii).

\smallskip\noindent
On montre maintenant~:\\
(4)\hspace{2em}$\force\pt q^P(q\neps{\cal Y}0\to q=p)$ \ et \
$\pt j\indi\pt q^P(q\neps{\cal Y}(j+1)\to\ex r^P(r\neps{\cal Y}j\land q=\psi(r,{\cal X}j)))$.\\
Pour cela, on définit le prédicat ${\cal Z}$ de SR$_1$, c'est-à-dire ${\cal Z}:\NN\fois P\to{\cal P}(\Pi)^{\NN}$,
en posant~:\\
${\cal Z}(0,q)(=\|q\neps{\cal Z}0\|)=\|q\neps{\cal Y}0\land q=p\|$~;\\
${\cal Z}(j+1,q)(=\|q\neps{\cal Z}(j+1)\|)=\|q\neps{\cal Y}(j+1)\land\ex r^P(r\neps{\cal Y}j\land q=\psi(r,{\cal X}j))\|$.\\
On a évidemment $\force q\neps{\cal Z}j\to q\neps{\cal Y}j$~; d'après (3), on a~:\\
$\force p\neps{\cal Z}0$ et $\force\pt j\indi\pt r^P(r\neps{\cal Z}j\to\psi(r,{\cal X}j)\neps{\cal Z}(j+1))$.\\
Par définition de ${\cal Y}$, on a donc $\force\pt n\indi\pt q^P(q\neps{\cal Y}n\to q\neps{\cal Z}n)$, d'où (4).

\smallskip\noindent
De la première partie de (4) et $\force p\neps {\cal Y}0$, \ on déduit (i).\\
De la seconde partie de (4), on déduit (iii), par récurrence sur $n$.\\
En appliquant le lemme~\ref{psi_decide}(iv), on en déduit également (v).

\smallskip\noindent
On montre (iv), par récurrence sur $m$. Le cas $m=0$ est conséquence de (iii). On suppose
$q\neps{\cal Y}n$, $r\neps{\cal Y}(n+m+1)$~; d'après (4), on a $r=\psi(r',{\cal X}(n+m))$. Par hypothèse
de récurrence, on a \ $r'\le q$. Le lemme~\ref{psi_decide}(ii) donne $r\le r'$, d'où $r\le q$.

\smallskip\noindent
On montre (vi) par récurrence sur $n$. Le cas $n=0$ est immédiat, d'après (i).
On suppose $q\neps{\cal Y}(n+1)$~; d'après (4), on a \ $q=\psi(q',{\cal X}n)$ avec $q'\neps{\cal Y}n$,
d'où $\C[q']$ par hypothèse de récurrence. Le lemme~\ref{psi_decide}(iii) donne alors $\C[q]$.

\cqfd

\smallskip\noindent
Le lemme~\ref{suite_cal_Y} montre que les hypothèses de la condition de chaîne dénombrable sont
vérifiées par le prédicat ${\cal Y}$. Si on pose \ $p^{\cal X}=\p({\cal Y})$, la conclusion de
la condition de chaîne donne \ $\force\pt n\indi\pt q^P(q\neps{\cal Y}n\to p^{\cal X}\le q)$ et
$\force\C[p]\to\C[p^{\cal X}]$.
Il en résulte que l'on a~:

\smallskip\noindent
$\force p^{\cal X}\le p$~; $\force\pt n\indi(p^{\cal X}\force^f\pm{\cal X}n)$ d'après
le lemme~\ref{suite_cal_Y}(v)~; \ $\force\C[p]\to\C[p^{\cal X}]$.\\
Cela termine la preuve du théorème~\ref{inf_chaine_den}.

\cqfd

\begin{theorem}\label{conserv_reels}
Si \ $\C$ a la propriété de chaîne dénombrable, il existe~:\\
une application $(p,{\cal X})\mapsto p^{\cal X}$ de $P\fois{\cal P}(\Pi\fois P)^{\NN}$ dans $P$~;\\
une application $(p,{\cal X})\mapsto K_{p,{\cal X}}$ de $P\fois{\cal P}(\Pi\fois P)^{\NN}$
dans ${\cal P}(\Pi)^{\NN}$~;\\
et deux quasi-preuves $\theta,\delta$ telles que~:\\
$(\theta,p^{\cal X})\fforce{\cal X}\simeq K_{p,{\cal X}}$ \ et \
$\delta\force\neg\C[pp^{\cal X}]\to\neg\C[p]$ \
quels que soient $p\in P$ et ${\cal X}\in{\cal P}(\Pi\fois P)^{\NN}$.
\end{theorem}\noindent
L'application $(p,{\cal X})\mapsto p^{\cal X}$ a été définie dans le théorème~\ref{inf_chaine_den}.
La quasi-preuve $\delta$ se déduit aisément de $\delta_0$ et $\delta_1$.\\
Pour $p\in P$ et ${\cal X}\in{\cal P}(\Pi\fois P)^{\NN}$, on définit alors
$ K_{p,{\cal X}}\in{\cal P}(\Pi)^{\NN}$ en posant~:\\
\centerline{$ K_{p,{\cal X}}(n)=\|p^{\cal X}\force^f{\cal X}n\|=
\|\pt q(\C[p^{\cal X}q]\to q\neps{\cal X}n)\|$.}

\begin{lemma}\label{theta01}
Il existe deux quasi-preuves \ $\theta_0,\theta_1$ telles que~:\\
$\theta_0\force(p^{\cal X}\force^f\pt n\indi( K_{p,{\cal X}}n\to{\cal X}n))$ et \
$\theta_1\force(p^{\cal X}\force^f\pt n\indi({\cal X}n\to K_{p,{\cal X}}n))$.
\end{lemma}\noindent
On a \ $p^{\cal X}\force^f( K_{p,{\cal X}}n\to{\cal X}n)\equiv
\pt q(q\force^f K_{p,{\cal X}}n\to p^{\cal X}q\force^f{\cal X}n)\equiv\\
\pt q((\C[q]\to K_{p,{\cal X}}n)\to p^{\cal X}q\force^f{\cal X}n)\equiv
\pt q((\C[q]\to p^{\cal X}\force^f{\cal X}n)\to p^{\cal X}q\force^f{\cal X}n)$.\\
Par définition de $\force^f$, cette formule s'écrit~:\\
$\pt q\pt r\{\C[q]\to\pt r((\C[p^{\cal X}r]\to r\neps{\cal X}n)),\C[p^{\cal X}qr]\to r\neps{\cal X}n\}$.\\
Elle est conséquence de $\C[p^{\cal X}qr]\to\C[p^{\cal X}r]\land\C[q]$~;\\
d'où une quasi-preuve $\theta_0$ telle que \ \
$\theta_0\force(p^{\cal X}\force^f\pt n\indi( K_{p,{\cal X}}n\to{\cal X}n))$.

\smallskip\noindent
On a \ $p^{\cal X}\force^f({\cal X}n\to K_{p,{\cal X}}n)\equiv
\pt q(q\force^f{\cal X}n\to p^{\cal X}q\force^f K_{p,{\cal X}}n)\equiv\\
\pt q(q\force^f{\cal X}n\to(\C[p^{\cal X}q]\to K_{p,{\cal X}}n))\equiv
\pt q((q\force^f{\cal X}n),\C[p^{\cal X}q]\to(p^{\cal X}\force^f{\cal X}n))$.\\
Or, cette formule n'est autre que $p^{\cal X}\force^f\pm{\cal X}n$. D'après le théorème~\ref{inf_chaine_den},
il existe donc une quasi-preuve $\theta_1$ telle que \
$\theta_1\force(p^{\cal X}\force^f\pt n\indi({\cal X}n\to K_{p,{\cal X}}n))$.

\cqfd

\smallskip\noindent
Preuve du théorème~\ref{conserv_reels}.\\
D'après le lemme~\ref{theta01} et le théorème~\ref{forcef-fforce}, on a~:\\
$(\chi_F\theta_0,p^{\cal X})\fforce\pt n\indi( K_{p,{\cal X}}n\to{\cal X}n)$ \ 
où $F$ est la formule \ $\pt n\indi(Yn\to X^+n)$~;\\
$(\chi_G\theta_1,p^{\cal X})\fforce\pt n\indi({\cal X}n\to K_{p,{\cal X}}n)$ \
où $F$ est la formule \ $\pt n\indi(X^+n\to Yn)$.

\cqfd

\begin{theorem}[Conservation des réels]\label{equiv_reels}\ \\
Si \ $\C$ a la propriété de chaîne dénombrable, il existe une quasi-preuve $\theta_0$ telle que~:\\
$(\theta_0,r)\fforce\pt X^+\ex X(X^+\simeq X)$ pour tout $r\in P$.
\end{theorem}\noindent
{\small{\bfseries Remarque.} Le sens intuitif est que, dans SR$_1$, les ensembles {\em d'entiers}
sont (extensionnellement) les mêmes que dans SR$_0$. Bien entendu, ce n'est pas le cas, en général,
pour les ensembles {\em d'individus}.}

\smallskip\noindent
On cherche $\theta_0$ tel que $(\theta_0,r)\fforce\pt X({\cal X}\simeq X\to\bot)\to\bot$ quel que soit
${\cal X}\in{\cal P}(\Pi\fois P)^{\NN}$.\\
Soit $(\xi,q)\fforce\pt X({\cal X}\simeq X\to\bot)$.
On veut avoir $(\theta_0,r)\star(\xi,q)\ps(\pi,q')\in\bbbot$, soit~:\\
$(\theta_0\star\xi\ps\pi,r(qq'))\in\bbot$ quels que soient $\pi\in\Pi$ et $r\in P$. Soit donc
$\tau\in\C[r(qq')]$~; on veut avoir~:\\
(1)\hspace{14em}$\theta_0\star\xi\ps\pi^\tau\in\bbot$.

\noindent
Par hypothèse sur $\xi$, on a $(\xi,q)\fforce{\cal X}\simeq K_{p,{\cal X}}\to\bot$ avec $p=r(qq')$.
Or, d'après le théorème~\ref{conserv_reels}, on a \
$(\theta,p^{\cal X})\fforce{\cal X}\simeq K_{p,{\cal X}}$. Il en résulte que \
$(\xi,q)\star(\theta,p^{\cal X})\ps(\varpi,\1)\in\bbbot$, soit~:\\
(2)\hspace{10em}$(\xi\star\theta\ps\varpi,q(p^{\cal X}\1))\in\bbbot$ pour toute $\varpi\in\Pi$.

\noindent
Soit $\alpha$ une quasi-preuve telle que, pour tout $\upsilon\in\C[(r(qq'))p']$, on ait \
$\alpha\upsilon\in\C[q(p'\1)]$.\\
On fixe $\upsilon\in\C[pp^{\cal X}]=\C[(r(qq'))p^{\cal X}]$, d'où $\alpha\upsilon\in\C[q(p^{\cal X}\1)]$.\\
D'après~(2), on a donc $\xi\star\theta\ps\varpi^{\alpha\upsilon}\in\bbot$, donc $\chi'\star\xi\ps\alpha\upsilon\ps\theta\ps\varpi\in\bbot$, pour toute $\varpi\in\Pi$.\\
On a donc montré \ $\lbd y((\chi'\xi)(\alpha)y)\theta\force\neg\C[pp^{\cal X}]$.\\
Or, d'après le théorème~\ref{conserv_reels}, on a une quasi-preuve \
$\delta\force\neg\C[pp^{\cal X}]\to\neg\C[p]$.\\
Comme $\tau\in\C[p)$, on en déduit \ $\delta\star\lbd y((\chi'\xi)(\alpha)y)\theta\ps\tau\ps\pi\in\bbot$.
On obtient donc le résultat (1) cherché en posant
$\theta_0=\lbd x(\chi)(\delta)\lbd y(\chi'x)((\alpha)y)\theta$.

\cqfd

\smallskip\noindent
Soit $F$ une formule restreinte de SR$_1$~; on définit, par récurrence, sa {\em traduction dans SR$_0$},
qui est une formule restreinte de SR$_0$, notée $F_0$~:\\
si $F$ est atomique de la forme $X(t_1,\ldots,t_n)$ ou $t\ne u$, alors $F_0=F$~;\\
si $F$ est atomique de la forme $X^+(t_1,\ldots,t_n)$, alors $F_0=X_0(t_1,\ldots,t_n)$, où
$X_0$ est une nouvelle variable de prédicat $n$-aire (dont le domaine est ${\cal P}(\Pi)^{\NN^n}$)~;\\
si $F=G\to H$, alors $F_0=G_0\to H_0$~;\\
si $F=\pt x\indi G$, alors $F_0=\pt x\indi G_0$~;\\
si $F=\pt X\,G$, alors $F_0=\pt X\,G_0$~;\\
si $F=\pt X^+G$, alors $F_0=\pt X_0\,G_0$.

\begin{theorem}\label{conserv_reels_2}\ \\
Si \ $\C$ a la propriété de chaîne dénombrable, alors pour toute formule $F[\ov{X}_1,\ldots,\ov{X}_n]$
close res\-treinte de SR$_1$, il existe deux quasi-preuves $\chi^+_F,\chi^-_F$ telles~que~:\\
$(\xi,p)\fforce F[\ov{X}_1,\ldots,\ov{X}_n]$ \ $\Fl$ \ $\chi^+_F\xi\force\C[p]\to F_0[X_1,\ldots,X_n]$~;\\
$\xi\force\C[p]\to  F_0[X_1,\ldots,X_n]$ \ $\Fl$ \ $(\chi^-_F\xi,p)\fforce F[\ov{X}_1,\ldots,\ov{X}_n]$~;\\
où $F_0$ est la traduction de $F$ dans SR$_0$.
\end{theorem}\noindent

\begin{lemma}\label{XsimeqY}
Soit $F[X_1,\ldots,X_n]$ une formule restreinte de SR$_0$. Alors, on a~:\\
$X_1\simeq Y_1,\ldots,X_n\simeq Y_n,F[X_1,\ldots,X_n]\;\vdash\;F[Y_1,\ldots,Y_n]$.
\end{lemma}\noindent
Immédiat, par récurrence sur $F$.

\cqfd

\smallskip\noindent
On montre alors le théorème~\ref{conserv_reels_2} par récurrence sur $F$. La preuve est la même que pour
le théorème~\ref{fforce-hat} lorsque $F$ est atomique, $F\equiv\pt X\,G$, $F\equiv\pt x\indi G$
ou $F\equiv(G\to H)$.\\
Il reste à examiner le cas où $F[X_1,\ldots,X_n]$ est de la forme $\pt X^+G[X,X_1,\ldots,X_n]$.\\
Si \ $(\xi,p)\fforce F[\ov{X}_1,\ldots,\ov{X}_n]$, on a \ $(\xi,p)\fforce G[\ov{X}_0,\ov{X}_1,\ldots,\ov{X}_n]$.
Par hypothèse de récurrence, on a \ $\chi_G^+\xi\force\C[p]\to G[\ov{X}_0,\ov{X}_1,\ldots,\ov{X}_n]$ et donc
\ $\chi_G^+\xi\force\C[p]\to\pt X_0G[X_0,\ov{X}_1,\ldots,\ov{X}_n]$.\\
On pose donc $\chi_F^+=\chi_G^+$.\\
Inversement, supposons que \ $\xi\force\C[p]\to\pt X_0G[X_0,\ov{X}_1,\ldots,\ov{X}_n]$~;
par hypothèse de récurrence, on a \ $(\chi^-_G\xi,p)\fforce G[\ov{X},\ov{X}_1,\ldots,\ov{X}_n]$
quel que soit le paramètre $\ov{X}$ de SR$_0$ et donc~:\\
(1)\hspace{8em}$(\chi^-_G\xi,p)\fforce\pt X\,G[X,\ov{X}_1,\ldots,\ov{X}_n]$.\\
Or, d'après le théorème~\ref{equiv_reels}, on a~:\\
(2)\hspace{8em}$(\theta_0,p)\fforce\pt X^+\ex X(X^+\simeq X)$.\\
Par ailleurs, d'après le lemme~\ref{XsimeqY}, on a~:\\
(3)\hspace{8em}$\vdash G[X,\ov{X}_1,\ldots,\ov{X}_n],X^+\simeq X\to G[X^+,\ov{X}_1,\ldots,\ov{X}_n]$.\\
Or (1,2,3) ont pour conséquence \ $\pt X^+G[X^+,\ov{X}_1,\ldots,\ov{X}_n]$. On déduit alors du
théorème~\ref{adequat_gen} (lemme d'adéquation), qu'il existe une quasi-preuve $\theta_1$ telle que~:\\
$(\theta_1,p)\fforce\pt X^+G[X^+,\ov{X}_1,\ldots,\ov{X}_n]$.

\cqfd

\section*{Un ultrafiltre sélectif sur $\mathbb{N}$}
On va utiliser une structure de réalisabilité construite avec la méthode décrite dans la section
\gmg Structures SR$_0$ et SR$_1$\gmd.
On considère donc d'abord un modèle de réalisabilité usuel, donné par un ensemble saturé
$\bbot\subset\Lbd_c\fois\Pi$. Les éléments de $\Lbd_c$ sont les $\lbd$-termes clos comportant
les continuations $\kk_\pi$, les instructions $\ccc,\sigma,\chi,\chi'$ et éventuellement d'autres.
Cette structure de réalisabilité sera désignée par SR$_0$.

\smallskip\noindent
Dans SR$_0$, on notera $\|F\|$ la valeur de vérité d'une formule close $F$ et \ $\xi\force F$
le fait que $\xi\in\Lbd_c$ réalise cette formule.\\
On construit maintenant une nouvelle structure de réalisabilité, désignée par SR$_1$. On prend pour
$P$ l'ensemble ${\cal P}(\Pi)^{\NN}$ muni de l'opération binaire $\land$, définie de la façon suivante~:\\
si $X,Y:\NN\to{\cal P}(\Pi)$ alors $(X\land Y)(n)=X(n)\land Y(n)$ pour tout $n\in\NN$ ($X(n),Y(n)\subset\Pi$
sont des valeurs de vérité de SR$_0$, donc $X(n)\land Y(n)=(X(n),Y(n)\to\bot)\to\bot$ est aussi une valeur
de vérité). L'élément distingué $\1$ de $P$ est la fonction telle que $\1(n)=\vide=\|\top\|$ pour tout
$n\in\NN$.

\smallskip\noindent
On a donc $\LLbd=\Lbd_c\fois{\cal P}(\Pi)^{\NN}$, $\PPi=\Pi\fois{\cal P}(\Pi)^{\NN}$ et
$\LLbd\star\PPi=(\Lbd_c\star\Pi)\fois{\cal P}(\Pi)^{\NN}$.

\smallskip\noindent
{\bfseries Notation.}
On utilisera les lettres $X,Y,U,V,\ldots$ pour désigner des variables de prédicat unaire dans SR$_0$
(leur domaine de variation est donc ${\cal P}(\Pi)^{\NN}=P$). Ce sont donc maintenant, en même temps,
des \emph{variables de condition}. Les lettres $A,B,\ldots$\ désigneront des prédicats particuliers
(éléments de ${\cal P}(\Pi)^{\NN}$).\\
Les variables de prédicat unaire de SR$_1$, variant dans
${\cal P}(\PPi)^{\NN}=({\cal P}(\Pi\fois{\cal P}(\Pi)^{\NN}))^{\NN}$ seront notées $X^+,Y^+,\ldots$\
Les constantes de prédicat unaire de SR$_1$ (éléments de ${\cal P}(\PPi)^{\NN}$) seront notées
${\cal X},{\cal Y},\ldots$

\smallskip\noindent
On considère la formule à une variable libre de prédicat unaire~:\\
\centerline{$\CC[X]\equiv\pt m\indi\ex n\indi X(m+n)$}

\noindent
qui exprime que $X\cap\NN$ est infini.\\
Rappelons que la notation $\pt m\indi F[m]$ représente la formule $\pt m($int$(m)\to F[m])$.\\
En utilisant le théorème~~\ref{memoire_ent}(ii) (mise en mémoire), on la remplacera ici par
$\pt m(\{s^m0\}\to F[m])$, puisqu'on la considère dans SR$_0$.\\
La formule \ $\CC[X]$ s'écrit donc \ $\pt m[\{s^m0\},\pt n(\{s^n0\},X(m+n)\to\bot)\to\bot]$.

\smallskip\noindent
{\small{\bfseries Remarque.} Nous utiliserons plus loin le quantificateur $\pt m^{\mbox{int}}$ dans SR$_1$. 
La notation $\pt m^{\mbox{int}}\,F[m]$ représentera alors la formule $\pt m(\{s^m0[]\}\to F[m])$,
c'est-à-dire $\pt m(\{(s^m0,\1)\}\to F[m])$.}

\smallskip\noindent
On termine la construction de la structure de réalisabilité SR$_1$ en définissant un sous-ensem\-ble
$\C[A]$ de $\Lambda_c$, pour tout $A\in{\cal P}(\Pi)^{\NN}$. On pose~:\\
\centerline{$\C[A]=\{\tau\in\Lbd_c;$ $\tau\force\CC[A]\}$.}

\noindent
On a donc \ $\bbbot=\{(\xi\star\pi,A);$ $\xi\in\Lbd_c,\pi\in\Pi,A:\NN\to{\cal P}(\Pi),
\pt\tau(\tau\force\CC[A]\Fl\xi\star\pi^\tau\in\bbot)\}$.

\smallskip\noindent
Dans SR$_1$, on notera $\vv F\vv$ la valeur de vérité d'une formule close $F$ (sous-ensemble de
$\PPi=\Pi\fois{\cal P}(\Pi)^{\NN}$) et \ $(\xi,X)\fforce F$ \ le fait que $(\xi,X)\in\LLbd=\Lbd_c\fois{\cal P}(\Pi)^{\NN}$
réalise cette formule.

\smallskip\noindent
Il faut noter qu'avec cette définition des l'ensembles $\C[p]$, on identifie les variables $p$ de condition
et les variables $X$ de prédicat unaire de SR$_0$. Dans les formules de SR$_0$, $\C[p]\to F$ sera
maintenant écrite $\CC[X]\to F$ et le quantificateur $\pt p^P$ est remplacé par $\pt X$.

\smallskip\noindent
On doit trouver des quasi-preuves $\alpha_i(0\le i\le4)$ telles que~:

\smallskip\noindent
$\alpha_0\force\pt X\pt Y\{\CC[X\land Y]\to\CC[X]\}$~; \
$\alpha_1\force\pt X\pt Y\{\CC[X\land Y]\to\CC[Y\land X]\}$~;\\
$\alpha_2\force\pt X\{\CC[X]\to\CC[X\land X]\}$~; \
$\alpha_3\force\pt X\pt Y\pt Z\{\CC[X\land(Y\land Z)]\to\CC[(X\land Y)\land Z)]\}$~;\\
$\alpha_4\force\pt X\{\CC[X]\to\CC[X\land\1]\}$.

\smallskip\noindent
Or, on a facilement~:\\
\centerline{$\vdash\HH:\pt x(Xx\to X'x)\to(\CC[X]\to\CC[X'])$ \ avec \
$\HH=\lbd f\lbd u\lbd m\lbd h(um)\lbd n\lbd x(hn)(f)x$.}\\
On peut donc poser $\alpha_i=\HH\beta_i$, pour des quasi-preuves $\beta_i(0\le i\le 4)$ telles que~:\\
$\beta_0\force\pt X\pt Y\{X\land Y\to X\}$~; \
$\beta_1\force\pt X\pt Y\{X\land Y\to Y\land X\}$~;\\
$\beta_2\force\pt X\{X\to X\land X\}$~; \
$\beta_3\force\pt X\pt Y\pt Z\{X\land(Y\land Z)\to(X\land Y)\land Z)\}$~;\\
$\beta_4\force\pt X\{X\to X\land\top\}$.\\
On peut donc poser $\beta_0=\lbd z(z)1$~; $\beta_1=\lbd z\lbd f((f)(z)0)(z)1$~;\\
$\beta_2=\beta_4=\lbd x\lbd f(f)xx$~; $\beta_3=\lbd z\lbd f((f)\lbd g((g)(z)1)(z)01)(z)00$\\
avec \ $1=\lbd x\lbd y\,x$ \ et \ $0=\lbd x\lbd y\,y$.

\subsection*{La condition de chaîne dénombrable}
On montre ici que la formule $\CC[X]$ satisfait la \emph{condition de chaîne dénombrable}.

\smallskip\noindent
Remarquons que le langage des formules de SR$_0$ contient maintenant trois relations d'équi\-valence
sur les prédicats unaires (dont le domaine de variation est ${\cal P}(\Pi)^{\NN}$).\\
Ce sont, en commençant par la plus forte~:\\
$\bullet$~~L'\emph{égalité de Leibniz}, que nous avons notée $X=Y$, et qui est l'égalité dans
l'ensemble $P$ de conditions. Elle est définie comme $\neg(X\ne Y)$~;
pour $X,Y\in{\cal P}(\Pi)^{\cal \NN}$, on a $\|X\ne Y\|=\Pi=\|\bot\|$ si $X$ et $Y$ sont la même
fonction, et $\|X\ne Y\|=\vide=\|\top\|$ si $X$ et $Y$ sont des fonctions différentes.\\
$\bullet$~~L'\emph{équivalence extensionnelle}, que nous avons notée $X\simeq Y$, qui est la formule~:\\
$\pt x\indi(Xx\dbfl Yx)$. Elle est associée à la relation de préordre $X\subseteq Y$ qui est la
formule~:\\
$\pt x\indi(Xx\fl Yx)$.\\
$\bullet$~~L'\emph{équivalence des conditions}, que nous avons notée $X\sim Y$, qui est la formule~:\\
$\pt Z(\CC[X\land Z]\dbfl\CC[Y\land Z])$. Elle équivaut à $\neg(\CC[X\setminus Y]\lor\CC[Y\setminus X])$.
Elle est associée à la relation de préordre $X\le Y$ qui est la formule \
$\pt Z(\CC[X\land Z]\to\CC[Y\land Z])$, qui équivaut à $\ex x\indi\pt y\indi(X(x+y)\to Y(x+y))$
ou encore $\neg\CC[X\setminus Y]$.\\
La fonction binaire $\setminus$ est définie sur $P={\cal P}(\Pi)^{\NN}$ par
$(X\setminus Y)(n)=\|X(n)\land\neg Y(n)\|$.

\smallskip\noindent
Il s'agit maintenant de construire une application $\p:{\cal P}(\Pi)^{\NN\fois P}\to P$
(avec $P={\cal P}(\Pi)^{\NN}$) et deux quasi-preuves ${\sf cd}_0,{\sf cd}_1$ telles que, pour tout
${\cal X}:\NN\fois P\to{\cal P}(\Pi)$, on ait~:\\
${\sf cd}_0\force\pt n\indi\pt Y(Y\neps{\cal X}n\to\p({\cal X})\le Y)~;$\\
${\sf cd}_1\force\pt n\indi\pt Y\pt Z(Y\neps{\cal X}n,Z\neps{\cal X}n\to Y=Z),\pt n\indi\ex Y(Y\neps{\cal X}n),
\pt n\indi\pt Y(Y\neps{\cal X}n\to\CC[Y]),\\
\hspace*{\fill}\pt m\indi\pt n\indi\pt Y\pt Z(Y\neps{\cal X}n,Z\neps{\cal X}(n+m)\to Z\le Y)\to\CC[\p({\cal X})]$.

\smallskip\noindent
On fixe donc ${\cal X}\in{\cal P}(\Pi)^{\NN\fois P}$ et on définit les formules~:\\
$A(i,n,{\cal X})\;\equiv\;\pt Y(Y\neps{\cal X}i\to Yn)$~; \
$B(j,n,{\cal X})\;\equiv\;\pt i\indi(i\le j\to A(i,n,{\cal X}))$~;\\
$\Phi(j,n,{\cal X})\;\equiv\;
j<n\land B(j,n,{\cal X})\land\pt k\indi(j<k<n\to\neg B(j,k,{\cal X}))$.\\
On définit alors la fonction $\p({\cal X}):\NN\to{\cal P}(\Pi)$ en posant \
$\p({\cal X})(n)=\|\ex j\indi\Phi(j,n,{\cal X})\|$.

\smallskip\noindent
{\small{\bfseries Remarque.}
Intuitivement, l'hypothèse de la condition de chaîne dénombrable exprime que ${\cal X}$ est une suite
d'ensembles à un seul élément, soit $\{A_i\}_{i\in\NN}$. La suite $A_i$ est une suite de parties de $\NN$,
qui est décroissante \emph{au sens des conditions}, c'est-à-dire que $A_i\setminus A_{i+1}$ est fini.
On définit une fonction partielle $f:\NN\to\NN$ en posant $f(j)=$ le premier entier $n\in\bigcap_{i\le j}A_i$ \
qui est ${>j}$, s'il existe. La relation $n=f(j)$ est définie par la formule $\Phi(j,n,{\cal X})$, et
$\p({\cal X})$ est défini comme l'image de $f$. Si chaque ensemble $A_i$ est une condition non triviale
(c'est-à-dire un ensemble infini d'entiers), alors l'image de $f$ est aussi infinie (et $f$ est totale).}

\smallskip\noindent
Pour obtenir la quasi-preuve ${\sf cd}_0$, il suffit de prouver la formule~:\\
\centerline{$\pt j\indi\pt Y(Y\neps{\cal X}j\to\p({\cal X})\le Y)$}

\noindent
à l'aide de la définition de $\p({\cal X})$ et de formules déjà réalisées par une quasi-preuve.

\smallskip\noindent
La formule \ $Y\neps{\cal X}j\to\p({\cal X})\le Y$ s'écrit \ $Y\neps{\cal X}j\to\neg\CC[\p({\cal X})\setminus Y]$,
c'est-à-dire~:\\
$Y\neps{\cal X}j\to\ex m\indi\pt{j'}\indi\pt l\indi[\Phi(j',m+l,{\cal X})\to Y(m+l)]$. On a donc~:\\
$\pt j\indi\ex m\indi\pt{j'}\indi\pt l\indi[\Phi(j',m+l,{\cal X})\to A(j,m+l,{\cal X})]\;\vdash\;
\pt j\indi\pt Y(Y\neps{\cal X}j\to\p({\cal X})\le Y)$\\
(parce que $\ex m\indi\pt Y(Y\neps{\cal X}j\to F)\;\vdash\;\pt Y(Y\neps{\cal X}j\to\ex m\indi F)$ quelle
que soit la formule $F$).\\
Il suffit donc de montrer la formule~:\\
\centerline{$\pt j\indi\ex m\indi\pt{j'}\indi\pt l\indi[\Phi(j',m+l,{\cal X})\to A(j,m+l,{\cal X})]$.}

\smallskip\noindent
{\bfseries Notation.} Pour alléger la notation, on écrira $m\le m'$ pour $\ex l\indi(m+l=m')$~; on supprimera l'exposant
\gmg int\gmdd des variables quantifiées et le para\-mètre ${\cal X}$ dans les formules $A,B$ et $\Phi$.\\
On écrira $(\pt Y\neps{\cal X}n)\,F$ pour $\pt Y(Y\neps{\cal X}n\to F)$ et $(\pt i\le j)\,F$ pour $\pt i(i\le j\to F)$.

\smallskip\noindent
On doit donc montrer~:\\
($\star$)\hspace{9em}$\pt j\ex m\pt j'\pt m'[\Phi(j',m'),m'>m\to A(j,m')]$.

\begin{lemma}\label{phi_jm_j'm'}\ \\
i) $\force\pt j\pt m\pt j'\pt m'[\Phi(j,m),\Phi(j',m'),m<m'\to A(j,m')\land j<j']$.\\
ii) $\force\pt j\pt j'\pt m'[\Phi(j',m'),j\le j'\to\ex m\,\Phi(j,m)]$.
\end{lemma}\noindent
On montre ces propositions en arithmétique du second ordre. On écrit~:\\
(1)~~$\Phi(j,m)\equiv j<m\land\pt i(i\le j\to A(i,m))\land\pt k(j<k<m\to\ex i(i\le j\land\neg A(i,k)))$~;\\
(2)~~$\Phi(j',m')\equiv j'<m'\land\pt i(i\le j'\to A(i,m'))\land\pt k(j'<k<m'\to\ex i(i\le j'\land\neg A(i,k)))$.\\
i) Si $j'\le j$, on a $j'<m$ d'après (1)~; on peut donc faire $k=m$ dans (2), puisque $m<m'$, ce qui donne \
$\ex i(i\le j'\land\neg A(i,m))$ donc $\ex i(i\le j\land\neg A(i,m))$, ce qui contredit (1)~; on a donc $j<j'$.
On peut alors faire $i=j$ dans (2), ce qui donne $A(j,m')$.\\
ii) De $\Phi(j',m')$ et $j\le j'$, on déduit $j<m'\land\pt i(i\le j\to A(i,m')$. Si $m$ est le premier entier
$m'$ qui a cette propriété, on a $\Phi(j,m)$.

\cqfd

\smallskip\noindent
On peut alors montrer $(\star)$~; l'entier $j$ étant fixé, il y a deux cas~:\\
$\bullet$~~Si on a $\ex m\,\Phi(j,m)$, le résultat découle immédiatement du lemme~\ref{phi_jm_j'm'}(i).\\
$\bullet$~~Sinon, d'après le lemme~\ref{phi_jm_j'm'}(ii), on a $\pt j'\pt m'[\Phi(j',m')\to j'<j]$. Si on a
$\pt j'\pt m'\neg\,\Phi(j',m')$, la propriété $(\star)$ est évidente. Sinon, soit $j_0$ le
plus grand entier tel que $\ex m\,\Phi(j_0,m)$. On a alors $\pt j'\pt m'[\Phi(j',m')\to j'\le j_0]$.
D'après le lemme~\ref{phi_jm_j'm'}(i), on a~:\\
$\pt j'\pt m\pt m'[\Phi(j_0,m),\Phi(j',m'),m< m'\to\bot]$, ce qui donne le résultat voulu, puisque l'on a
$\ex m\,\Phi(j_0,m)$.

\cqfd

\smallskip\noindent
Pour obtenir la quasi-preuve ${\sf cd}_1$, il suffit de prouver la formule \ $\pt j\ex m\,\Phi(j,m)$ à l'aide
des hypothèses. En effet, puisque qu'on a trivialement  \ $\vdash\Phi(j,m)\to j<m$, on en déduira~:\\
$\pt j\ex m(j<m\land\Phi(j,m))$, soit $\pt j\indi\ex m\indi(j<m\land\p({\cal X})(m))$ c'est-à-dire
$\CC[\p({\cal X})]$.

\begin{lemma}
$\vdash\pt j\pt m(j<m,B(j,m)\to\ex m\,\Phi(j,m))$.
\end{lemma}\noindent
Il suffit de considérer le premier entier $m$ tel que $j<m$ \ et \ $B(j,m)$.

\cqfd

\smallskip\noindent
Il reste donc à prouver la formule \ $\pt j\ex m(j<m\land B(j,m))$. On montre en fait la formule \
$\pt j\pt l\ex m(l\le m\land B(j,m))$~; ou encore~:

\smallskip\noindent
$(\star\star)$\hspace{7em}$\pt j\pt l(\ex m\ge l)(\pt i\le j)(\pt Y\neps{\cal X}i)Ym$.

\smallskip\noindent
On a les hypothèses~:\\
(H0)~~$\pt n\pt Y\pt Z(Y\neps{\cal X}n,Z\neps{\cal X}n\to Y=Z)$.\\
(H1)~~$\pt n\ex Y(Y\neps{\cal X}n$).\\
(H2)~~$\pt n\pt Y(Y\neps{\cal X}n\to\CC[Y])$.\\
(H3)~~$\pt j(\pt i\le j)\pt Y\pt Z(Z\neps{\cal X}i,Y\neps{\cal X}j\to\neg\CC[Y\setminus Z])$\\
(puisque $Y\le Z$ s'écrit $\neg\CC[Y\setminus Z]$).

\smallskip\noindent
Or, l'hypothèse (H3) donne~:\\
$\pt j(\pt i\le j)(\pt Z\neps{\cal X}i)(\pt Y\neps{\cal X}j)\ex l(\pt m\ge l)(Ym\to Zm)$.\\
Avec (H0), appliqué pour $n=j$, puis $n=i$, on en déduit (lemme~\ref{pt_ex_to_ex_pt}(ii))~:\\
$\pt j(\pt i\le j)\ex l(\pt Z\neps{\cal X}i)(\pt Y\neps{\cal X}j)(\pt m\ge l)(Ym\to Zm)$, soit~:\\
$\pt j(\pt i\le j)\ex l(\pt m\ge l)(\pt Y\neps{\cal X}j)(\pt Z\neps{\cal X}i)(Ym\to Zm)$, c'est-à-dire~:\\
$\pt j(\pt i\le j)\ex l(\pt m\ge l)(\pt Y\neps{\cal X}j))(Ym\to A(i,m))$,
puis (avec le lemme~\ref{pt_ex_to_ex_pt}(i))~:\\
$\pt j\ex l(\pt m\ge l)(\pt i\le j)(\pt Y\neps{\cal X}j))(Ym\to A(i,m))$ et enfin~:\\
$\pt j\ex l(\pt m\ge l)(\pt Y\neps{\cal X}j)(Ym\to B(j,m))$.\\
Or, la proposition $(\star\star)$ à démontrer s'écrit \ $\pt j\pt l(\ex m\ge l)B(j,m)$.\\
Grâce à l'hypothèse (H1), on est donc ramené à montrer \ $\pt j\pt l(\ex m\ge l)(\pt Y\neps{\cal X}j)Ym$,
ce qui est l'hypothèse~(H2).\\
On a utilisé~:
\begin{lemma}\label{pt_ex_to_ex_pt}\ \\
i)~~$(\pt i\le j)\ex l(\pt m\ge l)F\;\vdash\;\ex l(\pt m\ge l)(\pt i\le j)F$ \ si $l$ n'est pas libre dans $F$.\\
ii)~$\force(\pt Y\neps{\cal X}n)(\pt Z\neps{\cal X}n)(Y=Z),(\pt Y\neps{\cal X}n)\ex l\,F
\to\ex l(\pt Y\neps{\cal X}n)\,F$\\
quelle que soit la formule $F$.
\end{lemma}\noindent
i) Immédiat, puisque les quantificateurs sur les individus sont supposés restreints aux entiers.\\
ii) Conséquence de la formule réalisée $\pt Y\pt Z(Y=Z,G[Y]\to G[Z])$.

\cqfd

\subsection*{L'ultrafiltre}
{\bfseries Notations.}\\
Rappelons que $X\subseteq Y\;\equiv\;\pt n\indi(Xn\to Yn)$ ($X,Y$ sont des variables de prédicat unaires de SR$_0$ ou SR$_1$) et que $X\le Y\equiv\neg\CC[X\setminus Y]\;\equiv\;\ex l\indi\pt m\indi(l\le m,Xm\to Ym)$ lorsque $X$ et $Y$
sont des variables de condition, c'est-à-dire des variables de  prédicat unaire de SR$_0$.\\
On généralise cette notation aux variables de prédicat unaire de SR$_1$. On définit donc la formule de SR$_1$~: \
$X^+\le Y^+$ par $\ex l\indi\pt m\indi(l\le m,X^+m\to Y^+m)$.\\
Rappelons que, lorsque $F$ est une formule close de SR$_1$, on écrit \ $\fforce F$ pour exprimer qu'il existe
une quasi-preuve $\theta$ telle que $(\theta,\1)\fforce F$.

\smallskip\noindent
Dans la structure SR$_1$, on a défini le générique comme un prédicat unaire $J$ sur les conditions, en posant \
$\vv J(\ov{X})\vv=\Pi\fois\{\ov{X}\}$, \ pour tout $\ov{X}\in{\cal P}(\Pi)^{\NN}$. On définit, dans SR$_1$,
un prédicat ${\cal J}$ d'ordre~3, par la formule~:\\
\centerline{${\cal J}(X^+)\;\equiv\;\pt X(X\simeq X^+\to J(X))$}

\noindent
où $X,X^+$ sont des variables de prédicat unaire.

\begin{proposition}\label{JXsimeqJY}\ \\
i) $\fforce\pt X^+\pt Y^+(X^+\simeq Y^+,{\cal J}(X^+)\to{\cal J}(Y^+))$.\\
ii) $\fforce\pt X\pt X^+(X\simeq X^+\to(J(X)\dbfl{\cal J}(X^+)))$.
\end{proposition}\noindent
i) La formule considérée s'écrit \ $\pt X^+\pt Y^+\pt Y(X^+\simeq Y^+,{\cal J}(X^+),Y\simeq Y^+\to J(Y))$.\\
Elle est trivialement démontrable~: de $X^+\simeq Y^+,Y\simeq Y^+$, on déduit $Y\simeq X^+$~;\\
avec ${\cal J}(X^+)$, on en déduit $J(Y)$.\\
ii) Par définition de ${\cal J}$ , on a \ $\vdash X\simeq X^+,{\cal J}(X^+)\to J(X)$.\\
Inversement, d'après le théorème~\ref{elem_gen}(v), on a $\fforce\pt X\pt Y(J(X),Y\le X\to J(Y))$
et donc~:\\
$\fforce\pt X\pt Y(J(X),X\simeq Y\to J(Y))$. Or, cette formule a pour conséquence~:\\
$X\simeq X^+,J(X)\to\pt Y(Y\simeq X^+\to J(Y))$, c'est-à-dire $X\simeq X^+,J(X)\to{\cal J}(X^+)$.

\cqfd

\begin{theorem}\label{ultrafiltre1}
$\fforce\pt X^+\pt Y^+(Y^+\le X^+,{\cal J}(X^+)\to{\cal J}(Y^+))$.
\end{theorem}\noindent
D'après le théorème~\ref{elem_gen}(v), on a \ $\fforce\pt X\pt Y(Y\le X,J(X)\to J(Y))$.\\
Le théorème~\ref{equiv_reels} donne \ $\fforce\pt X^+\ex X(X^+\simeq X)$.\\
La proposition~\ref{JXsimeqJY}(ii) donne \ $\fforce\pt X\pt X^+(X\simeq X^+\to(J(X)\dbfl{\cal J}(X^+)))$.\\
Désignons ces trois formules par $F_0,F_1,F_2$ et soit $F$ la formule~:\\
$\pt X^+\pt Y^+(Y^+\le X^+,{\cal J}(X^+)\to{\cal J}(Y^+))$. On a~:\\
$(\theta_0,\1)\fforce F_0$~; $(\theta_1,\1)\fforce F_1$~; $(\theta_2,\1)\fforce F_2$ où
$\theta_0,\theta_1,\theta_2$ sont des quasi-preuves.\\
On a de plus \ $F_0,F_1,F_2\vdash F$ et donc $x_0:F_0,x_1:F_1,x_2:F_2\vdash t[x_0,x_1,x_2]:F$,\\
où $t$ est une quasi-preuve ayant les variables libres $x_0,x_1,x_2$.
D'après le théorème~\ref{adequat_gen} (lemme d'adéquation), on a donc
$t[(\theta_0,\1)/x_0,(\theta_1,\1)/x_1,(\theta_2,\1)/x_2]=
(t^*[\theta_0/x_0,\theta_1/x_1,\theta_2/x_2],\1)\fforce F$.

\cqfd

\begin{lemma}\label{gener_complm}
$\fforce\pt X(J(\1\setminus X)\dbfl\neg J(X))$.
\end{lemma}\noindent
D'après le théorème~\ref{elem_gen}(iv) et (vi), on a \
$\fforce\pt X(\neg J(X)\dbfl\pt Y(\neg\CC[X\land Y]\to J(Y)))$. Or, la formule
$\pt Y(\neg\CC[X\land Y]\to J(Y))$ équivaut à $\pt Y(Y\le(\1\setminus X)\to J(Y))$.
D'après le théorème~\ref{elem_gen}(v), on a \
$\fforce J(\1\setminus X)\dbfl\pt Y(Y\le(\1\setminus X)\to J(Y))$. On en déduit \
$\fforce\pt X(J(\1\setminus X)\dbfl\neg J(X))$.

\cqfd

\begin{theorem}\label{ultrafiltre2}\ \\
i) $\fforce\neg{\cal J}(\1)$.\\
ii) $\fforce\pt X^+({\cal J}(\1\setminus X^+)\dbfl\neg{\cal J}(X^+))$.\\
iii) $\fforce\pt X^+\pt Y^+[\neg{\cal J}(X^+),{\cal J}(X^+\land Y^+)\to{\cal J}(Y^+)]$.
\end{theorem}\noindent
i) D'après le théorème~\ref{elem_gen}(i), on a \ $\fforce\neg J(\1)$.
La proposition~\ref{JXsimeqJY}(ii) donne alors \ $\fforce\neg{\cal J}(\1)$.\\
ii) Le lemme~\ref{gener_complm} donne $\fforce\pt X(J(\1\setminus X)\dbfl\neg J(X))$. En appliquant la proposition~\ref{JXsimeqJY}(ii) et le théorème~\ref{equiv_reels}, on obtient le résultat voulu.\\
iii) D'après le théorème~\ref{elem_gen}(iii), on a \ $\fforce\pt X\pt Y[\neg J(X),J(X\land Y)\to J(Y)]$.
En appliquant la proposition~\ref{JXsimeqJY}(ii) et le théorème~\ref{equiv_reels}, on obtient
le résultat voulu.

\cqfd

\smallskip\noindent
{\small{\bfseries Remarque.} Les théorèmes~\ref{ultrafiltre1} et~\ref{ultrafiltre2} montrent que
${\cal J}$ est un idéal maximal non trivial sur l'anneau de Boole des parties de $\NN$.}

\subsection*{Programmes obtenus à partir de preuves}
Soit $F$ une formule close restreinte de l'arithmétique du second ordre (c'est-à-dire une formule du
second ordre dont tous les quantificateurs d'individu sont restreints à $\NN$).\\
Soit \ $x:\;$AU, $y:\;$ACD $\vdash t[x,y]:F$ \ une preuve de $F$ en arithmétique du second ordre, avec
l'axiome du choix dépendant ACD et l'axiome AU de l'ultrafiltre sur $\NN$, énoncé
sous la forme~: \gmg${\cal J}$ est un idéal maximal non trivial sur ${\cal P}(\NN)$\gmd.\\
On a donc $x:\,$AU $\vdash u[x]:G$ avec $u[x]=\lbd y\,t[x,y]$ et $G\equiv$ ACD $\to F$.\\
Dans la section précédente, on a obtenu une quasi-preuve $\theta$ telle que
$(\theta,\1)\fforce AU$.\\
Le théorème~\ref{adequat_gen} (lemme d'adéquation) donne $u[(\theta,\1)/x]\fforce G$ et donc
$(u^*[\theta/x],\1)\fforce G$.\\
D'après le théorème~\ref{conserv_reels_2}, on a donc $\chi^+_Gu^*[\theta/x]\force\CC[\1]\to G$,
c'est-à-dire~:\\
$\chi^+_Gu^*[\theta/x]\force\CC[\1],$ ACD $\to F$.
Or, on a des quasi-preuves $\xi_0\force\CC[\1]$ et $\eta_0\force$ ACD.\\
On a donc finalement $\chi^+_Gu^*[\theta/x]\xi_0\eta_0\force F$.\\
On peut alors appliquer au programme $\zeta=\chi^+_Gu^*[\theta/x]\xi_0\eta_0$ tous les résultats obtenus dans le cadre de la réalisabilité usuelle. Le cas où $F$ est une formule arithmétique (resp. $\Pi_1^1$) est étudié dans~[3]
(resp.~[4]).\\
Pour prendre un exemple très simple, si $F\equiv\pt X(X1,X0\to X1)$, on a
$\zeta\star\kappa\ps\kappa'\ps\pi\succ\kappa\star\pi$ quels que soient les $\lbd_c$-termes $\kappa,\kappa'$
et la pile~$\pi$.

\subsection*{Sélectivité}
On montre ici que l'ultrafiltre défini dans SR$_1$ par le prédicat $\neg{\cal J}$ est {\em sélectif} (voir [1]).
Cela signifie que, pour toute partition de $\NN$ dont aucun élément n'est dans l'ultrafiltre, il existe
un élément de l'ultrafiltre qui rencontre chaque élément de la partition en au plus un point.\\
{\bfseries Notation.} Dans cette section, tous les quantificateurs d'individu sont restreints aux entiers.
On écrira donc $\pt x$ au lieu de $\pt x\indi$.\\
$X,Y$ sont des variables de prédicat unaire, $Z$ une variable de prédicat binaire.\\
On désigne par~:\\
$\bullet$~~Part$[Z]$  la formule \ $\pt j\ex m\,Zmj\land\pt m\pt m'(Zmj,Zm'j\to m=m')$\\
(lire \gmg$Z$ est une partition de $\NN$\gmd).\\
$\bullet$~~Prem$[j,X,Z]$ la formule $\ex m[Xj\land Zmj\land(\pt i<j)\neg(Xi\land Zmi)]$\\
(lire \gmg $j$ est le premier élément de $X$ dans sa classe d'équivalence (mod. $Z$)\gmd).\\
$\bullet$~~R$[X,Z]$ la formule $\pt l\ex j(Xj\land(\pt m\le l)\neg Zmj)$\\
(lire \gmg $X$ rencontre une infinité de classes d'équivalence (mod. $Z$)\gmd).

\begin{lemma}\label{PartZptlexj}
Part$(Z)$, R$[X,Z]\;\vdash\;\pt n(\ex j>n)$Prem$[j,X,Z]$.
\end{lemma}\noindent
On a en effet~:\\
$\pt j\ex m\,Zmj$, R$[X,Z]\;\vdash\;\pt l(\ex m>l)\ex j(Xj\land Zmj)$\hspace{\fill}(1)\\
et, par ailleurs~:\\
$\pt m\pt m'(Zmj,Zm'j\to m=m')\;\vdash\;\pt n\ex l(\pt m>l)\pt j(Zmj\to j>n)$\hspace{\fill}(2)\\
On a donc~:\\
Part$(Z)$, R$[X,Z]\;\vdash\;\pt n\ex m[\ex j(Xj\land Zmj)\land\pt j(Zmj\to j>n)]$\hspace{\fill}(3)\\
(étant donné $n$, on choisit d'abord $l$ par (2), puis $m$ par (1)).\\
L'ordre sur $\NN$ étant bien fondé, on a \ $\vdash\pt m[\ex j\,F(j)\to\ex j(F(j)\land(\pt i<j)\neg F(i))]$
quelle que soit la formule $F$. Donc~:\\
$\vdash\pt m[\ex j(Xj\land Zmj)\to\ex j(Xj\land Zmj\land(\pt i<j)\neg(Xi\land Zmi))]$\hspace{\fill}(4)\\
De (3) et (4), on déduit~:\\
Part$(Z)$, R$[X,Z]\;\vdash\;\pt n\ex m(\ex j>n)[Xj\land Zmj\land(\pt i<j)\neg(Xi\land Zmi)]$\\
ce qui est le résultat voulu.

\cqfd

\begin{lemma}\label{negJ(X)ptY}
$\fforce\neg J(X),\pt Y\pt m(\pt j(Yj\dbfl Zmj)\to J(Y))\to$ R$[X,Z]$.
\end{lemma}\noindent
Il suffit de montrer que l'on a \
$\Gamma,\,\neg J(X),\,\pt Y\pt m(\pt j(Yj\dbfl Zmj)\to J(Y))\;\vdash$ R$[X,Z]$\\
où $\Gamma$ est la suite des propriétés du générique $J$ données au théorème~\ref{elem_gen} et au
lemme~\ref{gener_complm}.\\
La propriété (iii) du théorème~\ref{elem_gen} s'écrit \
$\fforce\pt Y\pt Y'(\neg J(Y),\neg J(Y')\to\neg J(Y\land Y'))$.\\
Le lemme~\ref{gener_complm} donne $\pt Y(J(\1\setminus Y)\dbfl\neg J(Y))$. On en déduit~:\\
\hspace*{13em}$\fforce\pt Y\pt Y'(J(Y),J(Y')\to J(Y\lor Y'))$\hspace{\fill}(5)\\
De $\pt Y\pt m(\pt j(Yj\dbfl Zmj)\to J(Y))$, on déduit donc, en raisonnant par récurrence sur $l$~:\\
\centerline{$\pt Y\pt l[\pt j(Yj\dbfl(\ex m\le l)Zmj)\to J(Y)]$.}\\
De $\neg J(X)$, on déduit $J(\1\setminus X)$ par le lemme~\ref{gener_complm}.\\
En appliquant (5) une fois de plus, on obtient donc~:\\
\centerline{$\pt Y\pt l[\pt j(Yj\dbfl\neg Xj\lor(\ex m\le l)Zmj)\to J(Y)]$.}\\
Or, le théorème~\ref{elem_gen}(i) donne $\fforce\neg J(\1)$. En faisant $Y=\1$ (c'est-à-dire $Yj\equiv\top$
pour tout $j$), on obtient \ $\pt l\ex j(Xj\land(\pt m\le l)\neg Zmj)$, c'est-à-dire R$[X,Z]$.

\cqfd

\smallskip\noindent
Désignons par Part$'(Z)$ la formule \ Part$(Z)\land\pt Y\pt m(\pt j(Yj\dbfl Zmj)\to J(Y))$\\
(lire \gmg$Z$ est une partition dont aucun élément n'est dans l'ultrafiltre\gmd).

\begin{lemma}\label{Part'ZptX}
$\fforce$ Part$'(Z),\pt X\pt Y(\pt j(Yj\dbfl$ Prem$[j,X,Z])\to J(Y))\to\bot$.
\end{lemma}\noindent
En appliquant les lemmes~\ref{PartZptlexj} and~\ref{negJ(X)ptY}, on obtient~:\\
$\fforce$ Part$'(Z)\to\pt X(\neg J(X)\to\pt n(\ex j>n)$Prem$[j,X,Z])$.

\smallskip\noindent
On applique alors le théorème~\ref{densite}, en définissant la fonction
$\phi:{\cal P}(\Pi)^{\NN}\to{\cal P}(\Pi)^{\NN}$ par~:\\
$\phi(X)(j)=\|$Prem$[j,X,Z]\|$ (noter que \ $\vdash$ Prem$[j,X,Z]\dbfl Xj\land$Prem$[j,X,Z]$).\\
On en déduit $\fforce$ Part$'(Z),\pt X\,J[\phi(X)]\to\bot$, ce qui est le résultat cherché.

\cqfd

\smallskip\noindent
Désignons par Part$^+(Z^+)$ la formule \ Part$(Z^+)\land\pt Y^+\pt m(\pt j(Y^+j\dbfl Z^+mj)\to{\cal J}(Y^+))$\\
(lire \gmg$Z^+$ est une partition dont aucun élément n'est dans l'ultrafiltre\gmd).

\begin{theorem}\label{selectif}\ \\
$\fforce\pt Z^+\{$Part$^+(Z^+),\pt X^+\pt Y^+(\pt j(Y^+j\dbfl$ Prem$[j,X^+,Z^+])\to{\cal J}(Y^+))\to\bot\}$.
\end{theorem}\noindent
Il suffit de remarquer que cette formule est conséquence de~:\\
$\pt Z\{$Part$'(Z),\pt X\pt Y(\pt j(Yj\dbfl$ Prem$[j,X,Z])\to J(Y))\to\bot\}$

(qui est donnée par le lemme~\ref{Part'ZptX})\\
$\pt X^+\ex X(X^+\simeq X)$, $\pt Z^+\ex Z(Z^+\simeq Z)$

(qui sont données par le théorème~\ref{equiv_reels})\\
et de $\fforce\pt X\pt X^+(X\simeq X^+\to(J(X)\dbfl{\cal J}(X^+)))$

(qui est donnée par la proposition~\ref{JXsimeqJY}(ii)).

\cqfd

\smallskip\noindent
Le théorème~\ref{selectif} exprime que, si $Z^+$ est une partition de $\NN$ dont aucun élément n'est
dans l'ultrafiltre, alors il existe une partie $X^+$ de $\NN\;$ et un élément $Y^+$ de l'ultrafiltre, tels que
$Y^+$ soit formé des éléments de $X^+$ qui sont les plus petits possibles dans leur classe d'équivalence.
Cela implique que $Y^+$ rencontre chaque classe d'équivalence en au plus un point. L'ultra\-filtre $\neg{\cal J}$
est donc sélectif.

\section*{Références}
[1]~~S. Grigorieff. {\em Combinatorics on ideals and forcing.}\\
Ann. Math. Logic 3(4) (1971), p. 363-394.

\smallskip\noindent
[2]~~J.-L. Krivine. {\em Typed lambda-calculus in classical Zermelo-Fraenkel set theory.}\\
Arch. Math. Log., 40, 3, p. 189-205 (2001).\\
http://www.pps.jussieu.fr/~krivine/articles/zf\_epsi.pdf

\smallskip\noindent
[3]~~J.-L. Krivine. {\em Dependent choice, `quote' and the clock.}\\
Th. Comp. Sc., 308, p. 259-276 (2003).\\
http://hal.archives-ouvertes.fr/hal-00154478\\
http://www.pps.jussieu.fr/~krivine/articles/quote.pdf

\smallskip\noindent
[4]~~J.-L. Krivine. {\em Realizability in classical logic.}\\
A paraître dans Panoramas et synthèses, Société Mathématique de France.\\
http://hal.archives-ouvertes.fr/hal-00154500\\
Version mise à jour à~:\\
http://www.pps.jussieu.fr/~krivine/articles/Luminy04.pdf

\smallskip\noindent
[5]~~J.-L. Krivine. {\em Realizability : a machine for Analysis and set theory.}\\
Geocal'06 (février 2006 - Marseille); Mathlogaps'07 (juin 2007 - Aussois).\\
http://cel.archives-ouvertes.fr/cel-00154509\\
Version mise à jour à~:\\
http://www.pps.jussieu.fr/~krivine/articles/Mathlog07.pdf

\end{document}